\documentclass[a4paper,11pt]{article}
\usepackage[DIV=12]{typearea}
\usepackage{longtable}
\usepackage{booktabs,colortbl}
\usepackage{multirow}
\usepackage{amsmath}
\usepackage{amssymb}
\usepackage{bbm}
\usepackage[utf8]{inputenc}
\usepackage{pdflscape}
\usepackage{xspace}
\usepackage[caption=false]{subfig}
\usepackage{nicefrac}
\usepackage{graphicx}
\usepackage{cite}  
\usepackage[small]{caption}
\usepackage{mathtools}
\usepackage{nccfoots}
\usepackage{tikz}
\usetikzlibrary{calc,positioning}
\usepackage[normalem]{ulem}

\usepackage[pdftex]{hyperref}
\hypersetup{
    pdftitle = {Spectral Distortions from Axion Monodromy Inflation},
    pdfauthor = {Henriquez-Ortiz, Mastache, Ramos-Sanchez}
}

\newcommand{\dd}{\mathrm{d}}

\advance \headheight by 3.0truept       % for 12pt mandatory...
\begin{document}

\begin{titlepage}

\vspace*{1.0cm}

\begin{center}
{\Large\textbf{\boldmath Spectral Distortions from 
Axion Monodromy Inflation\unboldmath}}

\vspace{1cm}
\textbf{Ra\'ul Henr\'iquez--Ortiz}$^{a,b}$,
\textbf{Jorge Mastache}$^{a,c}$, and
\textbf{Sa\'ul Ramos--S\'anchez}$^{d}$,
\Footnote{*}{%
raul.henriquez32@unach.mx; jh.mastache@mctp.mx; ramos@fisica.unam.mx
}
\\[5mm]
\textit{$^a$\small Facultad de Ciencias en F\'isica y Matem\'aticas, Universidad Aut\'onoma de Chiapas, \\ Carretera Emiliano Zapata Km.\ 8, Rancho San Francisco, Ciudad Universitaria, Ter\'an,\\ Tuxtla Guti\'errez, Chiapas, C.P.\ 29050, M\'exico}
\\[2mm]
\textit{$^b$\small Escuela de F\'isica, Facultad de Ciencias Naturales y Matem\'atica, Universidad de El Salvador,\\ final de Av.\ M\'artires y H\'eroes del 30 julio, San Salvador, C.P.\ 1101, El Salvador}
\\[2mm]
\textit{$^c$\small Consejo Nacional de Ciencia y Tecnolog\'ia, Av.\ Insurgentes Sur 1582, Col.\ Cr\'edito Constructor,\\ Alc.\ Benito Ju\'arez, C.P.\ 03940, M\'exico}
\\[2mm]
\textit{$^d$\small Instituto de F\'isica, Universidad Nacional Aut\'onoma de M\'exico, POB 20-364, Cd.Mx.\ 01000, M\'exico}
\end{center}

\vspace{1cm}

\vspace*{1cm}

\begin{abstract}
With the advent of new missions to probe spectral distortions of the cosmic microwave background with unprecedented precision, the study of theoretical predictions of these signals becomes a promising avenue to test our description of the early Universe. Meanwhile, axion monodromy still offers a viable framework to describe cosmic inflation. In order to explore new constraints on inflationary models based on axion monodromy while aiming at falsifying this scenario, we compute the spectral distortions predicted by this model, revealing oscillatory features that are compatible with Planck data. Further, the predicted distortions are up to 10\% larger than the signals obtained from the fiducial $\Lambda$CDM model and are observable in principle. However, contrasting with the predictions of the simplest power-law inflationary potentials challenges the falsifiability of axion monodromy as it would require to reduce at least 100 times the current forecast error of the PIXIE satellite, which shall be possible at some projected observational setups.
\end{abstract}

\end{titlepage}

\newpage

%%%%%%%%%%%%%%%%%%%%%%%%%%%%%%%%%%%%%%%%%%%%%%%%
\section{Introduction}
\label{sec:introduction}
%%%%%%%%%%%%%%%%%%%%%%%%%%%%%%%%%%%%%%%%%%%%%%%%

Cosmological inflation~\cite{Albrecht:1982wi,Guth:1980zm,Linde:1981mu} refers to the early epoch of exponential expansion of the Universe that sets up the initial conditions of the hot big bang.  It is perhaps the simplest scenario to address, among others, the flatness and horizon puzzles, as well as to explain the origin of the cosmic structure (see e.g.~\cite{Liddle:1999mq,Tsujikawa:2003jp,Vazquez:2018qdg, Achucarro:2022qrl} for some reviews). This process is triggered by a (pseudo-) scalar field known as the inflaton, whose potential is constrained to be sufficiently flat during a reasonable time frame, allowing for a rapid and accelerated expansion that flattens the Universe. Such a {\it slow-rolling} inflationary potential leads to a specific power spectrum of primordial (scalar and tensor) perturbations, the latter are eventually enhanced as the Universe expands, yielding 
the seeds for the large-scale structure that we observe today.
The observed scalar power spectrum of the Cosmic Microwave Background (CMB) is in general, described by the power law of the wavenumber $k$~\cite{Kosowsky:1995aa}
\begin{equation}
\label{eq:gralPowerSspectrum}
 P_{\mathcal R}(k) ~=~ \frac{2\pi^2}{k^3} \mathcal{A}_s\left(\frac{k}{k_\star}\right)^{n_s-1+\frac12\alpha_s(k)\ln{(\nicefrac{k}{k_\star})}}\,,
\end{equation}
where the amplitude $\mathcal A_s$ and the spectral index or tilt $n_s$ are observable parameters, $k_\star$ is the pivot scale of the experiment, and the tilt running $\alpha_s$ is a function that depends on the inflationary model.
One of the goals of current precision cosmology is to measure these quantities by, e.g., observing the polarization of the CMB \cite{CMB-S4:2016ple, Abazajian:2019eic, CMB-S4:2020lpa}. The ever-tighter constraints~\cite{POLARBEAR:2022dxa, Planck:2018vyg,BICEP:2021xfz, Tristram:2020wbi, SPIDER:2021ncy, PhysRevD.101.122003, Kusaka:2018yzq} on these observables and the ratio $r$ of the tensor and scalar amplitudes (associated with the gravitational wave background~\cite{Seljak:1996gy, Hu:1997hv, Kamionkowski:1997av}) offer an opportunity to discriminate among the various existing inflationary models (see e.g.~\cite{Martin:2013tda} for a large classification).

Inflationary models based on axion monodromy~\cite{Silverstein:2008sg,McAllister:2008hb}, on which we focus in this work, are motivated by the appearance in different scenarios of axions equipped with a low-energy potential that admits slow-roll over an extensive range in field space. For instance, in string compactifications, axions are particularly abundant~\cite{Svrcek:2006yi}. They arise, e.g., from dualizing gauge fields over nontrivial cycles in the compact space of a string compactification or from integrating $p$-forms along $p$-cycles in type II strings. Direct computations have shown that the canonically normalized fields associated with such axions exhibit in the large-field limit a monomial potential compatible with slow-roll~\cite{McAllister:2008hb,Pahud:2008ae,Flauger:2009ab}. These fields are super-Planckian but are naturally endowed with a sub-Planckian periodicity arising from an underlying shift symmetry. If worldsheet instantons or branes wrapping cycles are considered, this results in a small periodic modulation as a contribution to the low-energy effective potential of the axion~\cite{McAllister:2008hb}.
Similarly, from a purely bottom-up perspective, axion monodromy can arise from a scenario where couplings between an axion and a gauge field strength yield a monomial potential~\cite{Kaloper:2008fb} and gauge instantons produce the periodic modulation. Independently of its origin, besides providing a viable scope for large-field inflation, axion monodromy offers the appealing possibility of observable tensor modes~\cite{McAllister:2014mpa}.
The distinct features of axion monodromy further allow one to inspect other possible observables for its signals in the cosmological history of our Universe.

In the early Universe, photons and baryons are tightly coupled behaving as a single viscous fluid close to thermal equilibrium due to Compton, Bremsstrahlung, and double Compton scattering, processes that isotropize the photon-baryon fluid. However, early energy injection to the baryon-photon fluid can disrupt thermal equilibrium, causing the CMB to experience small departures from the blackbody distribution. These deviations are known as spectral distortions (SD) and are sensitive to any energy injected to the CMB at different epochs. CMB SD complement the anisotropy CMB observations and provide a new benchmark to test standard and non-standard cosmological scenarios at small scales. One mechanism in the canonical cosmological model that injects energy is Silk damping~\cite{Silk:1967kq}, a process that damps acoustic waves smaller than the sound horizon after the perturbation enters the horizon. Through this process, the energy stored at small scales in the radiation fluid is redistributed to larger scales resulting in SD~\cite{Daly:1991,Barrow:1991,Hu:1994bz,Cabass:2016giw,Chluba:2016bvg}. The thermal diffusion mechanism results in an increase in the average photon temperature. The emerging SD are the mixture of blackbody spectra from regions with different temperatures~\cite{Zeldovich1972, Chluba:2003exb}, and their amplitude is proportional to the square of the average temperature perturbations in the photon field ~\cite{Daly:1991, Zeldovich1969, Chluba:2012gq}. Therefore, the SD directly depends on the shape and amplitude of the primordial power spectrum of curvature perturbations, $P_{\mathcal R}(k)$. Many other standard processes can also contribute to produce SD, see e.g.~\cite{Hu:1994bz,Khatri:2012tv,Chluba:2013wsa,Chluba:2013dna,Chluba:2019kpb,Rubino-Martin:2006hng}. There are also various non-standard mechanisms that induce SD, such as decay and annihilation of relic particles, evaporation of primordial black holes, primordial magnetic fields, and cosmic strings, see e.g.~\cite{Diacoumis:2017hff,Chluba:2019nxa,Chluba:2013pya,Lucca:2019rxf}.
The damping of modes over the wavenumber interval $50~{\rm Mpc}^{-1} \lesssim k \lesssim 10^4~{\rm Mpc}^{-1}$ (equivalent to redshifts $5\times 10^4 \lesssim z \lesssim 2\times 10^6 $) dissipate their energy while creating a non-zero chemical potential creating $\mu$ SD, and the damping modes with $k < 50~{\rm Mpc}^{-1}$ ($z \lesssim 10^4$) results in a so-called $y$ SD, also related to the (thermal and kinematic) Sunyaev-Zeldovich (SZ) effect~\cite{Zeldovich1969}.

SD were tightly constrained in 1996 by COBE/FIRAS to be $|\mu| \leq  9 \times 10^{-5}$ and $|y|  \leq 1.5 \times 10^{-5}$ ($95\%$ C.L.)~ \cite{Fixsen:1996nj}. Since models predict typically smaller SD, greater experimental precision had to be awaited to resume research on SD. Fortunately, new experimental missions, such as the Primordial Inflation Explorer (PIXIE) and its enhanced version Super-PIXIE will soon explore SD with expected standard errors $\sigma(\mu) \simeq 3\times10^{-8}, \sigma(y) \simeq 3.4\times10^{-9}$~\cite{Kogut:2011xw} and $\sigma(\mu) \simeq 7.7\times10^{-9}, \sigma(y) \simeq 1.6\times10^{-9}$~\cite{Chluba:2019nxa, Delabrouille:2019thj}, respectively. Moreover, there even exist proposals of alternative configurations of PIXIE with 1000 times improved sensitivity, to achieve $\sigma(\mu) = 1.5\times10^{-11}$ and $\sigma(y) = 1.2\times10^{-12}$ ($68\%$ C.L.)~\cite{Fu:2020wkq}. Such sensitivity will be instrumental in falsifying inflationary models. We are hence driven to provide precise descriptions of the SD features of inflationary models. Recent efforts in this direction have been done for various models~\cite{Chluba:2012we,Cabass:2016giw,Chluba:2016bvg,Cho:2017zkj,Bae:2017tll,Schoneberg:2020nyg}, but we still lack the predictions of axion monodromy, which we aim at providing in this work.

We organize this work as follows. In section~\ref{sec:framework} a brief review of the main properties of inflationary models based on axion monodromy is presented, followed in section~\ref{sec:sd} by an overview of theory of CMB SD for primordial small-scale perturbations. Using these elements, in section~\ref{sec:sdam} we compute and discuss the potentially observable features of SD in axion-monodromy inflation, which are the main result of our investigation. In section~\ref{sec:ns-r}, we discuss the compatibility of axion-monodromy inflation with current measurements of $r$ and $n_s$. Finally, in section~\ref{sec:conclusion} a brief discussion of our results and outlook is given.
 
%%%%%%%%%%%%%%%%%%%%%%%%%%%%%%%%%%%%%%%%%%%%%%%%
\section{Axion monodromy}
\label{sec:framework}
%%%%%%%%%%%%%%%%%%%%%%%%%%%%%%%%%%%%%%%%%%%%%%%%

In inflationary models based on axion monodromy, the potential energy of a canonically normalized axion $\phi$ can be written as
\begin{equation}
\label{eq:potential}
  V(\phi) ~=~ V_0(\phi) + \Lambda^{4} \cos\left( \frac{\phi}{f}+\gamma_0 \right)\,,
\end{equation}
where $\gamma_0$ is a phase that is usually ignored (even though it naturally appears in this scenario and has an important impact on the value of $n_s$, as we discuss in section~\ref{sec:ns-r}).
Here, the amplitude of the periodic modulation is given by the scale $\Lambda$ related to the strong dynamics that give rise to the potential, and $f$ is the dimensionful decay parameter of the axion $\phi$. In string constructions that support slow-roll inflation, it is known that in the large-field regime, the potential $V_0(\phi)$ can adopt the monomial structure~\cite{Easther:2013kla,Flauger:2014ana}
\begin{equation}
    V_0(\phi) ~\approx~ \lambda^{4-p}\phi^p
    \qquad\text{with}\quad p < 2
\end{equation}
and a scale $\lambda$, which can be fixed by the amplitude $\mathcal A_s$ of the scalar primordial fluctuations. Clearly, $V_0(\phi)\neq0$ breaks the axion shift symmetry $\phi \to \phi+2\pi f$ and  induces a monodromy, as the potential changes after each period. The potential~\eqref{eq:potential} has been shown to be useful to describe cosmological inflation, especially in scenarios with large tensor-to-scalar ratios~\cite{McAllister:2014mpa} (where recent observations favor small over large powers, see section~\ref{sec:ns-r}).
In this case, since $\phi$ can interact during inflation with moduli before these achieve their stabilization, both the oscillation amplitude and the axion decay parameter vary in general, i.e.\ $\Lambda^4=\Lambda^4(\phi)$ and $f=f(\phi)$. These dependencies induce a drift in the amplitude and in the oscillations, which may leave an imprint on cosmological observations. In some string models, the frequency drift can be expressed (at leading order in the slow-roll parameters) as~\cite{Flauger:2014ana}
\begin{equation}
 f(\phi) ~=~ f_0 \left(\frac{\phi}{\phi_\star}\right)^{-p_f}\,,
\end{equation}
where $f_0$ is the standard axion decay constant, $p_f$ is a drift parameter that encodes the dynamics of $\phi$ and moduli, and is adopted to be of order unity. Further, $\phi_\star$ is the value of the field when the pivot scale $k_\star$ exits the horizon, i.e.~such that for a fixed $k_\star$ the relation $k_\star=a(\phi_\star) H(\phi_\star)$ is satisfied. We focus here on the main signature of models with axion monodromy, which is the oscillatory behavior, considering that the periodic modulation is a small (nonperturbative) contribution. Hence, we ignore the drift in $\Lambda$ and assume that the modulation depends on a small parameter $b\ll 1$ that relates $\Lambda$ with the scale $\lambda$  according to~\cite{Flauger:2010ja}
\begin{equation}
  b ~:=~ \frac{\Lambda^4}{V_0'(\phi_\star)f(\phi_\star)}
    ~=~ \frac{\Lambda^4}{\lambda^{4-p}p\,\phi_\star^{p-1}f_0}\,.
\end{equation}

With these ingredients, using perturbation theory, one can solve the background equation of motion for $\phi$ in the slow-roll regime and approximately linear potential, and then compute via the Mukhanov-Sasaki equation the primordial scalar power spectrum at leading order in $b$ and in the limit\footnote{Our units are such that the reduced Planck mass is the unity, $M_\mathrm{Pl}=1$.} $f_0 \phi_\star \ll p $, ideal for observable non-Gaussianities~\cite{Flauger:2010ja}. The primordial scalar power spectrum reads~\cite{Flauger:2014ana,Easther:2013kla}
\begin{equation}
\label{eq:pkmndy1} 
 P_{\mathcal{R}}(k) = \frac{2\pi^2}{k^3} \mathcal{A} _{s} \left( \frac{k}{k_{\star}}\right)^{n_{s}-1} \left\lbrace 1 
+ \delta n_{s}  \cos \left[  \frac{\phi_\star}{f_0} \left( \frac{\phi_{k}}{\phi_\star}\right)^{p_{f}+1} + \vartheta \right]  \right\rbrace \,,
\end{equation}
where $\phi_k$ denotes the value of the field when its mode with wavenumber $k$ exits the horizon, which for slow-roll conditions is approximately given by
\begin{equation}
\label{eq:phik}
\phi_{k} ~\approx~ \sqrt{2p \left(N_0-\ln{\left( \tfrac{k}{k_{\star}}\right) }\right) }
\qquad\text{with}\qquad 
N_0~:=~N_\star+\phi_\mathrm{end}^2/2p\,.
\end{equation}
Here, $N_{\star}$ is the number of e-folds, $\phi_\mathrm{end}\approx p/\sqrt2$ corresponds to the value of the field at the end of slow-roll inflation, which can be determined by solving when any of the slow-roll parameters,
\begin{equation}
\label{eq:slow-roll}
    \epsilon_V ~:=~\frac12\left(\frac{V'}{V}\right)^2
    \approx~\frac12\left(\frac{V'_0}{V_0}\right)^2,\qquad 
    \eta_V~:=~\frac{V''}{V}
    \approx~\frac{V''_0}{V_0} 
    \qquad\text{with}\qquad
    V'~:=~\frac{dV}{d\phi}\,,
\end{equation}
violates the conditions $\epsilon_V,|\eta_V|\ll1$. We use the condition $\epsilon_V(\phi_\mathrm{end})=1$. Thus, from eq.~\eqref{eq:phik} we see that $\phi_\star \approx \sqrt{2pN_0}$. Further, the amplitude of the oscillatory contribution to the power spectrum is given in this limit by
\begin{equation}
\label{eq:pkmndy2}
\delta n_{s} ~=~ 3\,b\,\sqrt{ \frac{2\pi}{\alpha}}
\qquad\text{with}\qquad
\alpha ~:=~ (p_f+1)\frac{\phi_\star}{2f_0 N_0}\left(\frac{\sqrt{2p N_0}}{\phi_\star} \right)^{p_f+1}\!\approx~
\frac{p\,(p_f+1)}{f_0\,\phi_\star} \,.
\end{equation}
Note that small $b$ and $f_0\phi_\star\ll p$ implies $\delta n_s\ll1$ too. Finally, we find that in this limit 
\begin{equation}
    \label{eq:as2v}
    \mathcal A_s ~=~\left.\frac{V}{24\pi^2 \epsilon_V}\right|_{\phi=\phi_\star}
    \approx~\left.\frac{V_0}{24\pi^2 \epsilon_{V_0}}\right|_{\phi=\phi_\star}
    =~\frac{\lambda^{4-p}}{12\pi^2 p^2}\phi_\star^{p+2}\,,
\end{equation}
which allows us to fit $\lambda$ for different values of $p$ by comparing with Planck's best fit value $\ln(10^{10}\mathcal A_s)=3.0448$~\cite[Table 1]{Planck:2018vyg}.  In table~\ref{tab:ModelParameters}, we list the values of $\lambda,\phi_\star$ and $\phi_\mathrm{end}$ for three benchmark choices of $p$ with fixed $k_\star=0.05\,\mathrm{Mpc}^{-1}$ and $N_\star =57.5$. 

\begin{table}[t!]
    \centering
    \begin{tabular}{c|ccc|cc}
    $p$           & $\phi_\star$ & $\phi_\mathrm{end}$ & $10^4 \lambda$ & $n_s$ & $r$\\
    \hline
    $\nicefrac23$ &   $8.77$ & $0.47$ & $14.4$ & $0.977$ & $0.046$\\
    $1$           &  $10.75$ & $0.71$ & $5.85$ & $0.974$ & $0.069$\\
    $\nicefrac43$ &  $12.42$ & $0.94$ & $1.78$ & $0.971$ & $0.092$
    \end{tabular}
    \caption{Approximate values of some parameters for our benchmark choices of $p$ in inflationary models based on axion monodromy. We consider $k_\star=0.05\,\mathrm{Mpc}^{-1}$ and $N_\star=57.5$.  We take here the approximation $V\approx V_0\sim\phi^p$. Both values for the inflaton field $\phi$ and the scale $\lambda$ are given in units of $M_{\rm Pl}$. For completeness, we provide $n_s$ and $r$, which have been computed using eq.~\eqref{eq:nsandr} in this approximation. The tension of these results against observable data is discussed in section~\ref{sec:ns-r}.}
    \label{tab:ModelParameters}
\end{table}

As we are interested in the compatibility with observations of inflation based on axion monodromy, it is useful to recall that for a model of inflation characterized by the potential $V(\phi)$, the values of the scalar tilt $n_s$ and the tensor-to-scalar ratio $r$ are respectively given in terms of the slow-roll parameters by
\begin{equation}
\label{eq:nsandr}
  n_s ~=~ 1-6\epsilon_V(\phi_\star)+2\eta_V(\phi_\star)
  \qquad\text{and}\qquad
  r ~=~ 16\epsilon_V(\phi_\star)\,,
\end{equation} 
where $\epsilon_V$ and $\eta_V$ are defined in eq.~\eqref{eq:slow-roll}.

%%%%%%%%%%%%%%%%%%%%%%%%%%%%
%%% SPECTRAL DISTORTIONS %%%
\section{Spectral distortions}
\label{sec:sd}
%%%%%%%%%%%%%%%%%%%%%%%%%%%%
Spectral distortions are classified depending on their spectral shape and are directly related to the thermodynamic history of the photons. At $z \gtrsim 2 \times 10^6$, Compton scattering is the dominant collision process; it is efficient, driving any perturbation to thermodynamic equilibrium and maintaining the spectrum of the CMB close to a black body. In the range $5\times 10^4 \lesssim  z \lesssim 2\times 10^6$, number-density changing processes, such as Bremsstrahlung and double Compton scattering, become inefficient due to the expansion of the Universe, resulting in a Bose-Einstein distribution with a non-zero chemical potential $\mu$ for the photons which is approximately constant, even though it is in general a function of frequency and time~\cite{Tashiro:2014pga}. For redshifts satisfying\footnote{This classical picture has been recently refined, and it is now understood that there is a gradual transition from $\mu$ to $y$ at redshifts in the range $5\times 10^{4} \lesssim  z \lesssim 3 \times 10^{5}$. The information on the transition is carried out in the residual $r$-type distortions~\cite{Chluba:2013pya,Chluba:2013wsa}.  We will not consider this intermediate epoch and assume the transition between the $\mu$ and $y$ eras to be instantaneous at a redshift $z \approx 5 \times 10^4$.} $z  \lesssim 5 \times 10^4$, Compton scattering becomes inefficient, and background electrons at a higher temperature can boost the CMB photons out of equilibrium to create $y$ SD. The latter can be regarded as an analog of the Sunyaev–Zeldovich effect for the early Universe~\cite{SunyaevZel1969}.

Perturbation theory in a cosmological framework aims at understanding the underlying physics of the CMB. When the perturbations enter the horizon, they enter as acoustic waves and are immediately affected by thermal conductivity and viscosity. This process, known as Silk damping~\cite{Silk:1967kq}, damps waves smaller than the sound horizon. This wave damping releases energy, which is redistributed to the radiation bath, increasing the photon temperature and producing SD~\cite{SunyaevZel1970} by mixing blackbody spectra with different temperatures~\cite{Chluba:2012we, Daly:1991}. The magnitude of $\mu$ and $y$ SD are proportional to the square of the amplitude of the waves that are damped~\cite{Daly:1991}, which is contained in the primordial power spectrum, governed by cosmic inflation. 

Observable SD are expected to be on scales smaller than galaxies, restricting the amplitude of initial perturbations on small scales and complementing the results from CMB anisotropy observations which are constrained at small scales by Silk damping. This leads to constraints on the initial power spectrum at high wavenumbers, $1 \lesssim k \lesssim 10^4 \; \mathrm{Mpc}^{-1}$.
Precise measurements of the CMB power spectrum along with SD constraints from COBE/FIRAS observations~\cite{Mather:1993ij, Fixsen:1996nj} established upper bounds for $\mu \lesssim 9\times 10^{-5}$ and $y \lesssim 1.5 \times 10^{-5}$ at $2\sigma$ C.L. (see also milder bounds by ARCADE 2~\cite{Arcade2_experiment}). The TRIS experiment~\cite{Gervasi:2008eb} found the slightly stronger constraint $\mu \lesssim 6\times 10^{-5}$ for frequencies close to $\nu \simeq 1$ GHz. The absence of primordial black holes and ultracompact minihalos put upper bounds on the small-scale primordial power spectrum amplitude. The latter put constraints on the amplitude of the primordial power spectrum less than $0.01-0.06$ over for the wavenumber range $0.01 \lesssim k \lesssim 10^{23} \; {\rm Mpc^{-1}}$. Forthcoming concepts such as PIXIE and its extensions~\cite{Chluba:2019nxa,Kogut:2019vqh} will be able to explore previously inaccessible SD scales, furthering our understanding of the inflationary epoch. This demands revisiting the status of the theoretical predictions.

The total distortion of the photon intensity spectrum, $\Delta I$, can be written analytically (at first order) in terms of the $\Delta T$ temperature shift, $y$ and $\mu$ distortions as follows~\cite{Lucca:2019rxf}
\begin{equation}\label{eq:sd1}
\Delta I (\nu)
~\approx~ \frac{ \Delta T}{T} G (\nu) + y\, Y_{\mathrm{SZ}} (\nu)  + \mu\, M_{\mathrm{SZ}} (\nu)\,.
\end{equation}
The different terms can be determined by using the Green’s function approach introduced in~\cite{Fu:2020wkq}. The first term is proportional to the function
\begin{equation}
    G(\nu)~=~T \,\frac{\partial B}{\partial T}  \,,
    \qquad\text{with}\quad
    B(\nu)~:=~\frac{2 h \nu^{3}}{c^{2}\left(e^{x}-1\right)}
    \quad\text{and}\quad
    x~:=~\frac{h \nu}{k_{B} T}\,.  
\end{equation}
The spectral shape-functions multiplying $y$ and $\mu$ parametrize the out-of-equilibrium Compton effect and chemical potential~\cite{Zeldovich:1969ff}, and are defined as
\begin{subequations}
\begin{eqnarray}
    Y_{\mathrm{SZ}}(\nu) &\simeq& G(\nu)(x \operatorname{coth}(x)-4)\,, \\
    M_{\mathrm{SZ}}(\nu) &\simeq& G(\nu)\left(0.4561 - \frac{1}{x} \right)\,. 
\end{eqnarray}
\end{subequations}

Note that the contribution $\Delta T\, G(\nu)/T$ in eq.~\eqref{eq:sd1} arises from $B(T+\Delta T) - B(T)$ and, hence, does not distort the blackbody spectrum. It corresponds just to  a departure of the average temperature of the CMB with respect to the blackbody temperature today, difference that is besides very difficult to observe~\cite{Lucca:2019rxf}. Therefore, we shall focus here on $y$ and $\mu$ SD.

A good estimate of $\Delta I$ can be achieved by computing the components of the effective energy release caused by the dissipation of primordial acoustic modes. This yields~\cite{Chluba:2013dna}
\begin{subequations}\label{eq:sd2}
\begin{eqnarray}
  y &\approx& \frac{1}{4} \int_{0}^{z_{\mu,y}} \dd z \frac{1}{a^{4}\rho_{\gamma}} \frac{\dd (a^{4} Q_{ac})}{\dd z} \,, \\
  \mu &\approx& 1.4 \int^{\infty}_{z_{\mu,y}} \dd z  \frac{\mathcal{J}_{bb}(z)}{a^{4}\rho_{\gamma}} \frac{\dd (a^{4} Q_{ac})}{\dd z} \,,
\end{eqnarray}
\end{subequations}
where $z_{\mu,y} \approx 5 \times 10^{4}$, $\mathcal{J}_{bb}(z)\approx \exp{-[z/z_\mu]^{5/2}}$  parametrizes
the thermalization efficiency and accounts for the effects of photon
production/annihilation by double Compton scattering, and the factor $a^{-4}\rho_{\gamma}^{-1} \dd(a^{4} Q_{ac})/\dd z$ quantifies the energy release caused by the dissipation of primordial acoustic modes and encodes all the evolution of the radiation field for some initial conditions~\cite{Chluba:2012gq,Chluba:2013dna}.

An analytic approximation for the effective rate of energy release arising from the damping of adiabatic modes is given by (see~\cite{Fu:2020wkq} for details)
\begin{equation}\label{eq:sd4}
     \frac{1}{a^{4}\rho_{\gamma}} \frac{\dd (a^{4} Q_{ac})}{\dd z} ~=~ 2 C^{2}  \int_{k_\mathrm{min}}^{\infty} \frac{k^{4} \dd k}{2 \pi^{2}} P_{\mathcal{R}}(k) \partial_z k^{-2}_{D} e^{-2k^{2}/k^2_D} \,.
\end{equation}
Note that the appearance of the primordial power spectrum $P_\mathcal{R}(k)$, as defined in eq.~\eqref{eq:gralPowerSspectrum}, emphasizes an unavoidable bond between SD and inflation, as the structure of $P_\mathcal{R}(k)$ is defined by the inflationary process.
For adiabatic modes $C^2\approx(1+\nicefrac{4R_\nu}{15})^{-2}\approx0.813$ 
is a normalization coefficient with $\nicefrac{4R_\nu}{15}$ accounting for the correction due to the anisotropic stress in the neutrino fluid.
Furthermore, $k_D \approx 4.048 \times10^{-6}(1+z)^{\nicefrac32}$ ${\rm Mpc}^{-1}$ is the photon damping scale. 
Inserting eq.~\eqref{eq:sd4} in eq.~\eqref{eq:sd2} and performing the integration over $z$, one arrives at a simple integral expression for $y$ and $\mu$ SD~\cite{Chluba:2013dna}
\begin{equation}
\label{eq:sd5}
i ~\approx~ \int_{k_\mathrm{min}}^{\infty} \frac{k^{2} \dd k}{2 \pi^2} P_{\mathcal{R}}(k)\, W^{i} (k)\,,
\end{equation}
where $k_\mathrm{min} =  1~\text{Mpc}^{-1}$ and $W^i(k)$ are the so-called $k$-space window functions for the SD of type  $i\in\{y,\mu\}$, which account for the acoustic damping and thermalization effects, given by
\begin{subequations}
\begin{eqnarray}
W^{y}(k) &\approx& 
\frac{C^{2}}{2} \,\mathrm{e}^{-2 k^{2} / k_{\mathrm{D}}^{2}\left(z_{\mu, y}\right)} 
\approx \frac{C^{2}}{2} \,\mathrm{e}^{-k^{2} / 32^2}\,,\\
W^{\mu} (k) ~&\approx&~ 2.8\, C^{2}  \exp{ \left(  -\frac{\left[ \frac{\hat{k}}{1360} \right]^{2} }{1 + \left[ \frac{\hat{k}}{260} \right]^{0.3} + \left[ \frac{\hat{k}}{340} \right]} \right)} - 5.6\,  W^{y}(k)\,.
\end{eqnarray}
\end{subequations}
Here we introduced the dimensionless wavenumber $\hat{k} := \nicefrac{k}{k_\mathrm{min}}$. This approximation is valid when the underlying model is similar to the concordance model, which applies in particular for models based on axion monodromy with parameters that comply with the constraints imposed by Planck data. 

Previous works have shown that SD can place robust bounds on the amplitude of the primordial power spectrum arising from various inflationary potentials (exhibiting peculiar features, such as bumps, kinks, and discontinuities)~\cite{Chluba:2013wsa,Chluba:2013dna,Chluba:2012we,Chluba:2012gq}. In the following we focus on the study of the SD produced by inflationary models based on axion monodromy.

%%%%%%%%%%%%%%%%%%%%%%%%
\section{Spectral distortions from axion monodromy} \label{sec:sdam}
%%%%%%%%%%%%%%%%%%%%%%%%

\begin{figure}[t!]
\begin{center}
\includegraphics[width=0.5\textwidth]{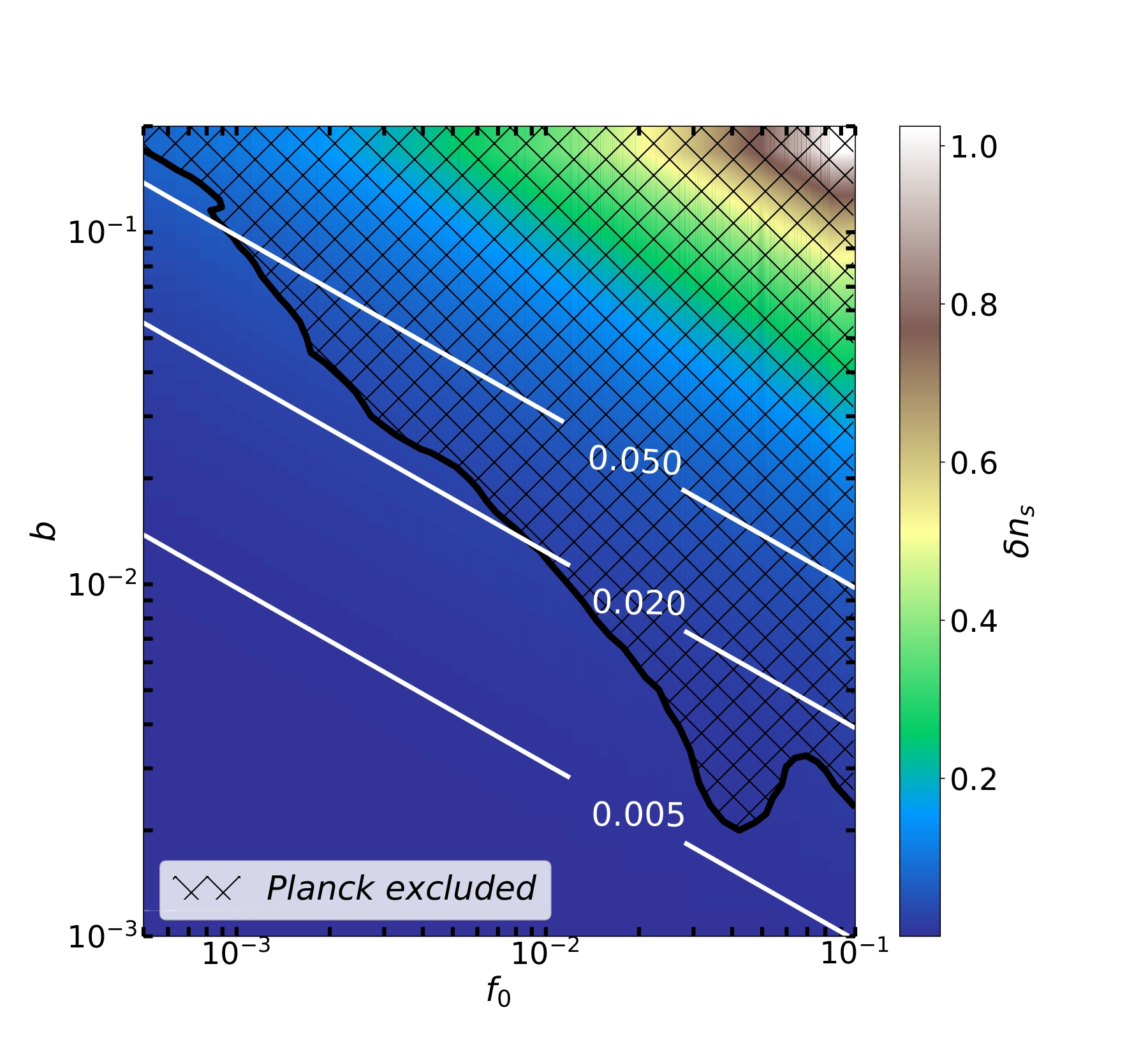}
\caption{Constraints at 68\% C.L.\ for the axion decay constant $f_0$ and the modulating parameter $b$ for $p=\nicefrac43$ based on Planck's data~\cite[fig.~32]{Planck:2018jri}. The shaded region has been excluded. The heatmap refers to values of $\delta n_s$ associated with $(b,f_0)$ according to eq.~\eqref{eq:pkmndy2} with our choice of $p_f=-0.7$, $N_\star=57.5$ and $\phi_\star=12.38\ M_\textrm{Pl}$. Some of the $\delta n_s$ values are marked along the white lines.}
\label{plot:axm1}
\end{center}
\end{figure}

We aim now to determine the observational features of $\mu$ and $y$ SD in the inflationary scenario based on axion monodromy discussed in section~\ref{sec:framework}, whose features could soon be accessible to experiments. With this goal in mind, we compute the magnitudes of these SD associated with different choices of the main parameters of our model. 

The free parameters of our model are: the modulation $b$, the axion decay constant $f_0$, the monomial power  $p$, the oscillation drifting power $p_f$, and the phase $\vartheta$, which we set to zero for simplicity. To compare with observations and previous results, we set the pivot scale to $k_\star=0.05\,$Mpc$^{-1}$, and the number of e-folds at that scale to $N_\star=57.5$, assuming instantaneous reheating. As discussed in section~\ref{sec:framework}, $\phi_\star$, $\phi_\mathrm{end}$ and $\lambda$ can be determined from the previous free parameters and the observed value of $\mathcal A_s$, and $\delta n_s$
is calculated via eq.~\eqref{eq:pkmndy2}. For the benchmark values of $p\in\{\nicefrac23,1,\nicefrac43\}$, we find the values displayed in table~\ref{tab:ModelParameters}.

For different choices of the free parameters, we compute $\mu$ and $y$ SD, eq.~\eqref{eq:sd5}, using the primordial power spectrum for axion monodromy, eq.~\eqref{eq:pkmndy1}.
The parameter space that we explore is bounded to the following values
\begin{equation}
\label{eq:paramSpace}
10^{-3} \leq b \leq 0.1\,,\qquad 
9 \times 10^{-4}\leq f_0\leq 0.01\,,\qquad\text{and}\qquad 
-0.75 \leq p_f\leq 1.0\,.    
\end{equation}
We adopt these values because they favor sizable SD while complying with all our priors. In addition, as we will shortly discuss, this selection is found within the parameter window explored by Planck.

\begin{figure}[t!]
\begin{center}
\includegraphics[width=0.325\textwidth]{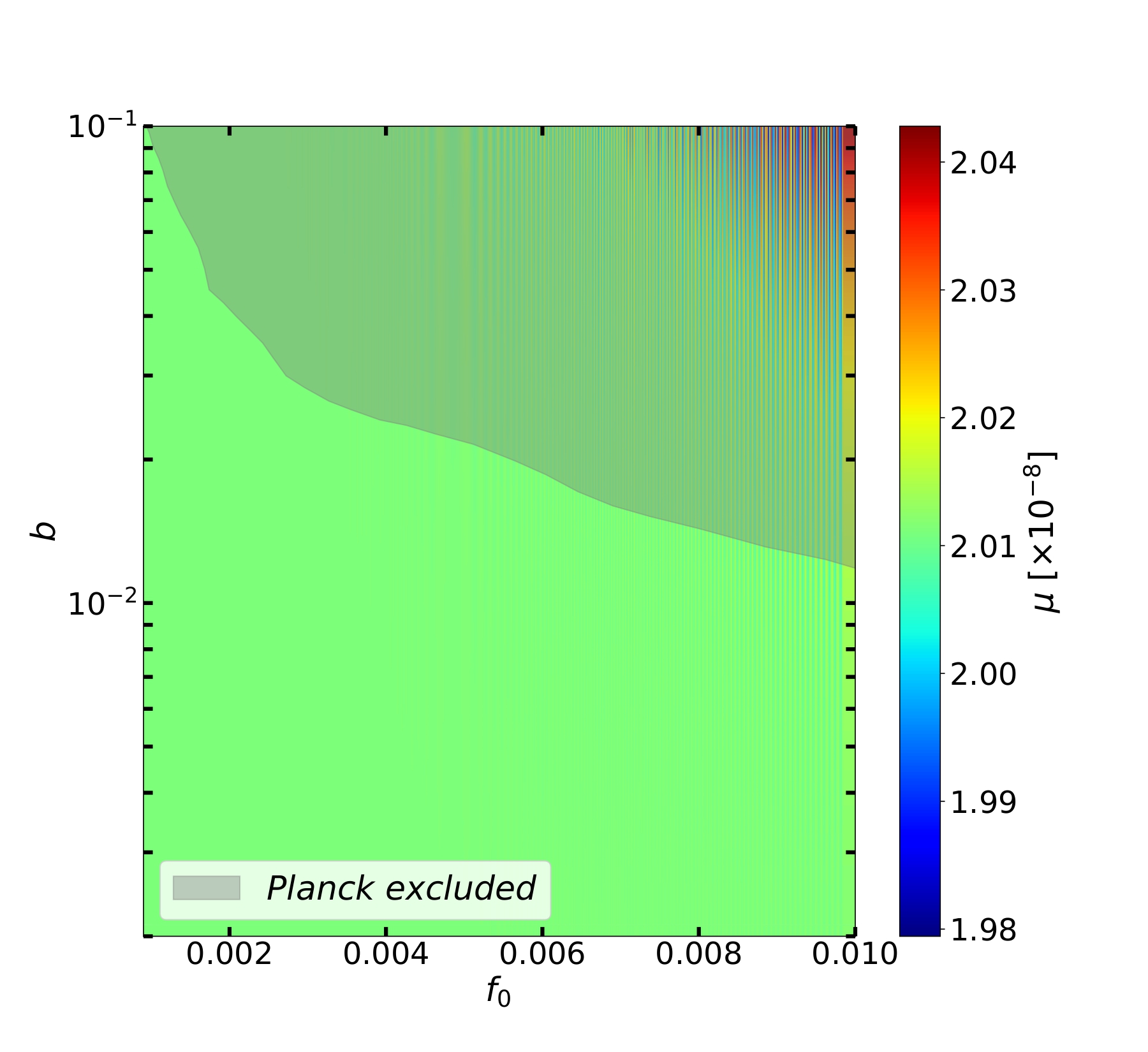}
\includegraphics[width=0.325\textwidth]{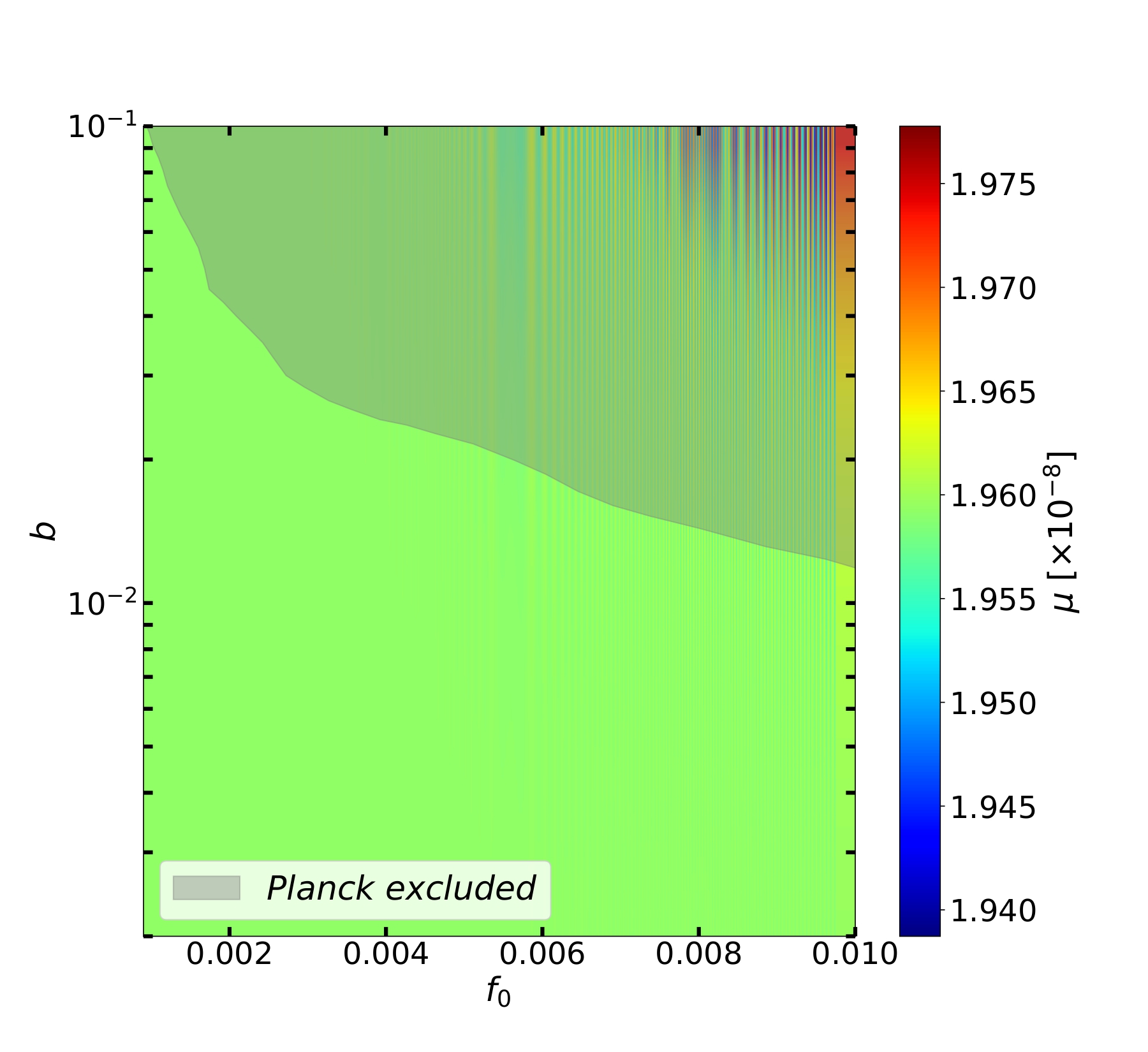} 
\includegraphics[width=0.325\textwidth]{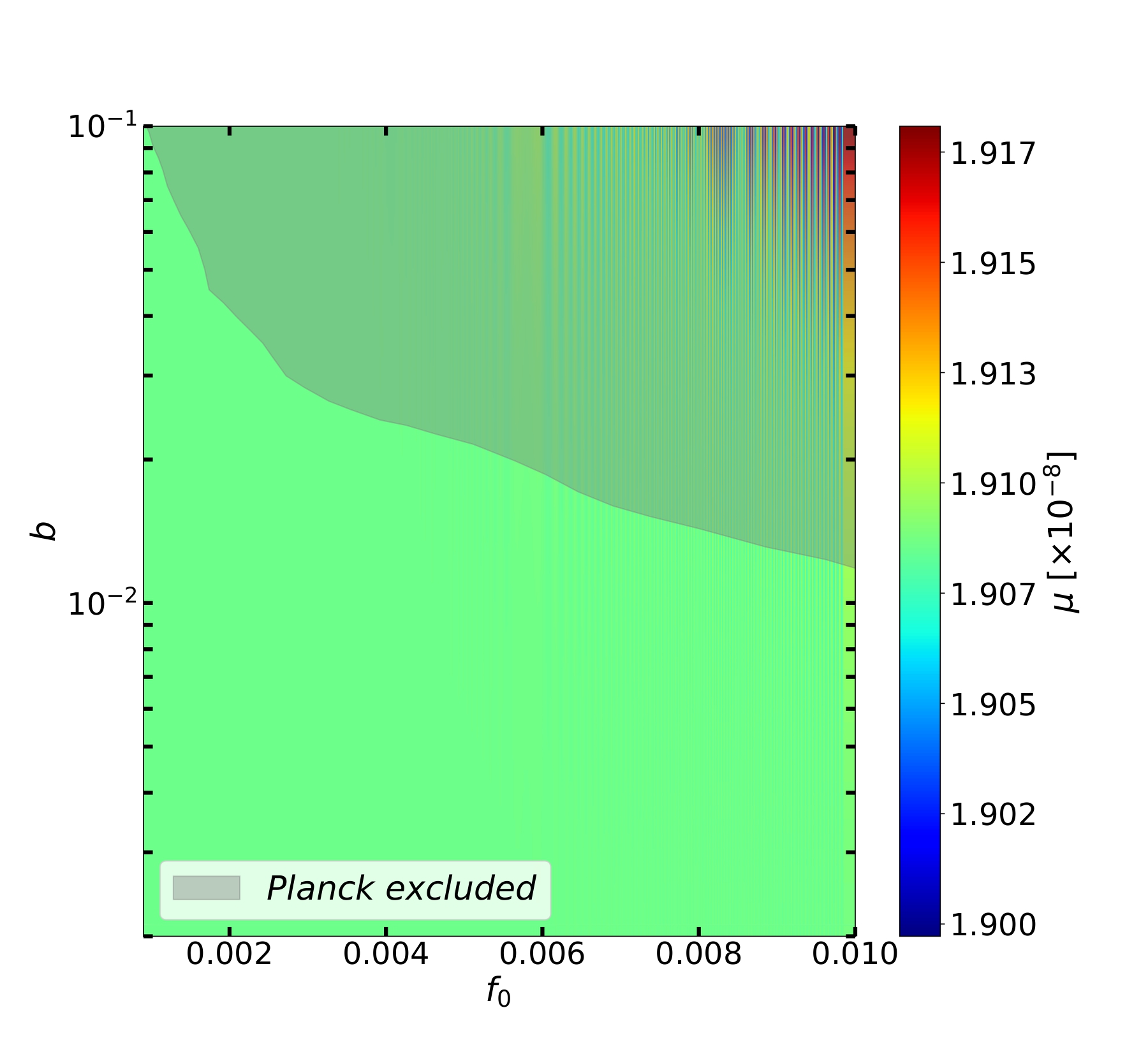}
\includegraphics[width=0.325\textwidth]{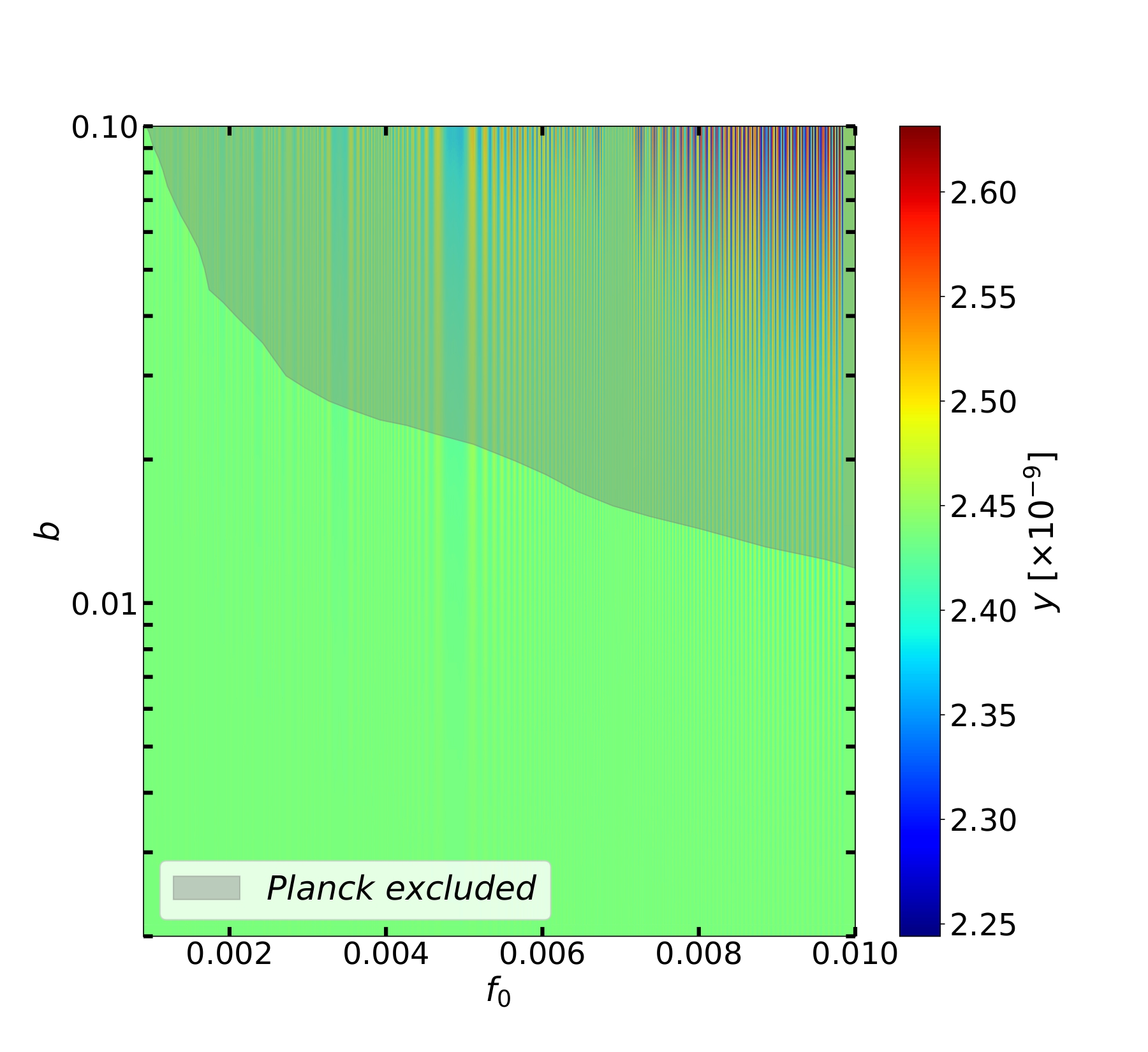}
\includegraphics[width=0.325\textwidth]{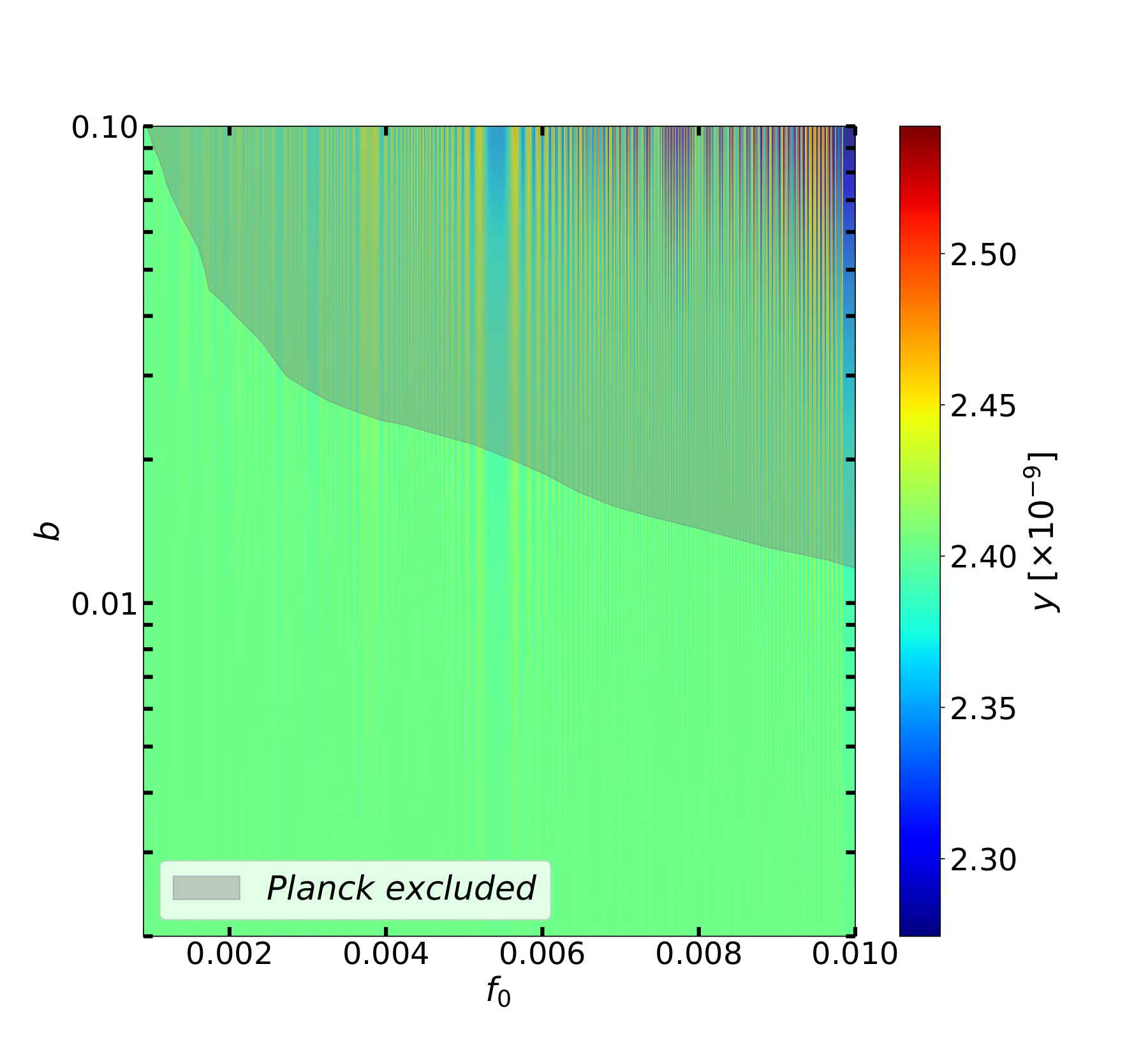}
\includegraphics[width=0.325\textwidth]{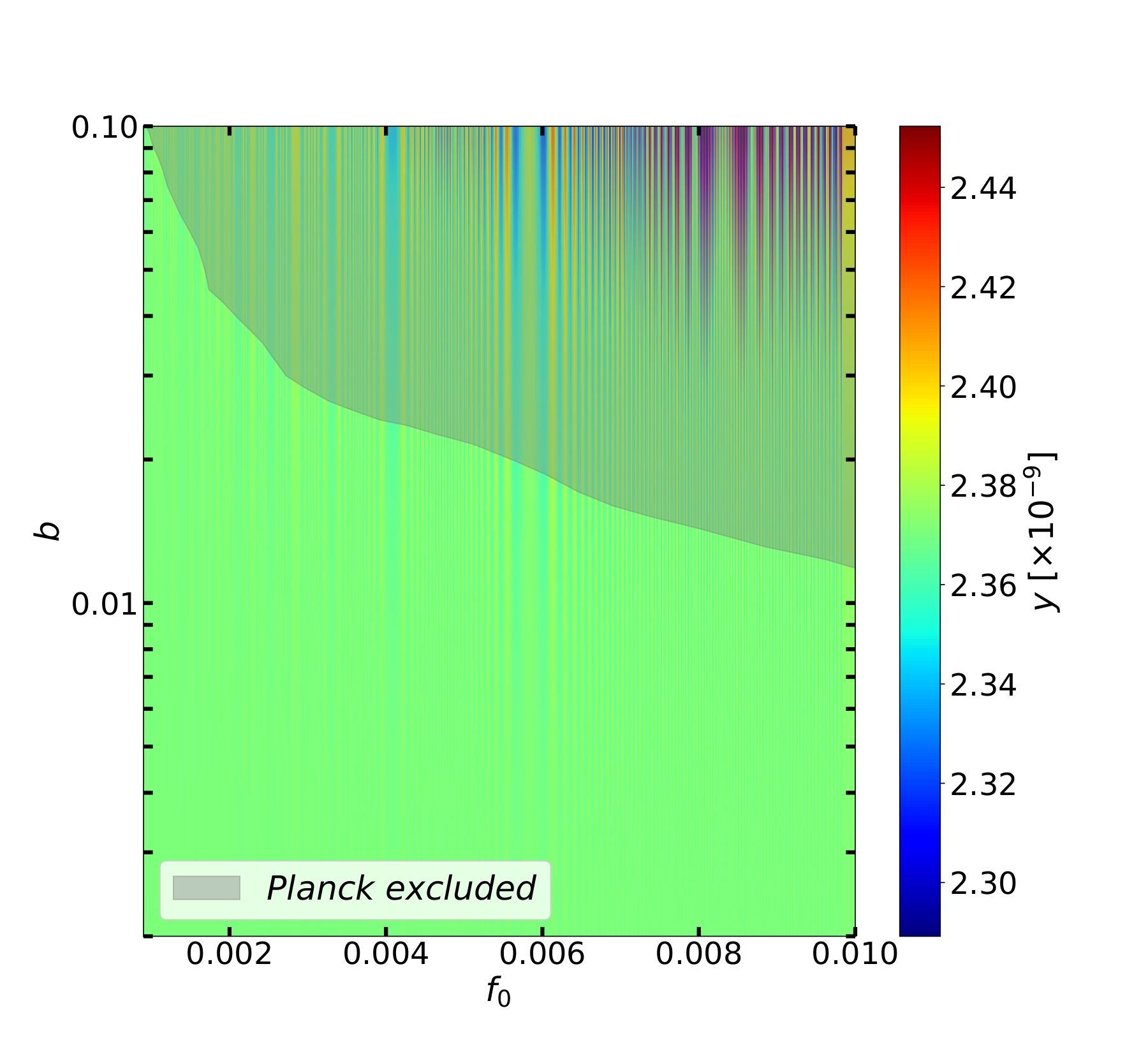}
\caption{$\mu$ (top) and $y$ (bottom) SD in axion-monodromy inflation, computed via eq.~\eqref{eq:sd1} in the $(b,f_0)$ parameter space. We explore here the region with $10^{-3}\leq b \leq 0.1$ and $9 \times 10^{-4}\leq f_0\leq 0.01$. We take our three benchmark values $p=\nicefrac23$ (left), $1$ (middle), $p=\nicefrac43$ (right), and fixed $p_f =-0.7$. The gray-shaded area corresponds to the region excluded by Planck at 68\% C.L.\ as shown in figure~\ref{plot:axm1}. All plots are very similar since most of the (colored) structure lies in the excluded region. There are some differences: the admissible (green) regions lead to the various predictions of table~\ref{tab:dominant-predictions}. Also, we observe some smaller allowed values of $y$ (blue stripes) for $0.005<f_0<0.006$.}
\label{plot:axm2}
\end{center}
\end{figure}

To achieve compatibility with observations we subject the parameters to the bounds set by the Planck collaboration~\cite[section 7.4]{Planck:2018jri}. Planck established limits on $f_0$, $p_f$ and $\delta n_s$ in axion monodromy models with $p\in\{\nicefrac23,1,\nicefrac43\}$. We translate through the relation~\eqref{eq:pkmndy2} the bounds of $f_0$ and $\delta n_s$ to constraints on $f_0$ as function of the modulation parameter $b$, for fixed $p$ and $p_f$. Taking $p=\nicefrac43$ and $p_f=-0.7$ to maximize these bounds, we show in figure~\ref{plot:axm1} the $(b,f_0)$ region of parameter space that is consistent with Planck data at 68\% C.L. We see that the admissible values lead to small $\delta n_s$, which justifies the blue color in the heatmap scale. From~\cite[fig.~32]{Planck:2018jri}, we read off that the values of $p_f$ in our parameter space~\eqref{eq:paramSpace} are all allowed if the combination of $(b,f_0)$ is chosen to comply with figure~\ref{plot:axm1}. The constraints for $p=\nicefrac23$ and $1$ are similar but somewhat milder~\cite{Planck:2018jri}. Hence, using the restrictions of figure~\ref{plot:axm1} for all our choices of $p$ leads to a conservative scenario, where results based on these constraints are compatible with observations.

In figure~\ref{plot:axm2} we display the predictions of inflationary models based on axion monodromy for $\mu$ (top) and $y$ (bottom) SD as functions of the modulation amplitude $b$ and the axion decay constant $f_0$, for fixed $p_f=-0.7$ and different values of $p=\nicefrac23$ (left), $1$ (middle) and $\nicefrac43$ (right). We fix $p_f=-0.7$ because this value delivers the largest SD, as we shall shortly see. We shade in gray the region in parameter space excluded by Planck, according to the bounds presented in figure~\ref{plot:axm1}. The predominantly predicted values of SD (appearing in green) are given in table~\ref{tab:dominant-predictions}. By inspecting the heatmap values in the figure, we note that arbitrary $b$ and $f_0$ only lead to SD values that differ by up to 1\% with respect to the dominant predictions.

\begin{table}[t!]
    \centering
\begin{tabular}{c|ccc}
                     & $p=\nicefrac23$ & $p=1$   & $p=\nicefrac43$\\
    \hline
     $10^8\mu$ & $2.011$         & $1.956$ & $1.908$ \\
     $10^9y$   & $2.438$         & $2.404$ & $2.371$    
\end{tabular}
    \caption{Dominant predicted values of $\mu$ and $y$ SD by inflationary models with axion monodromy for a choice monomial powers $p$ and fixed $p_f=-0.7$.
    \label{tab:dominant-predictions}}
\end{table}

\begin{figure}[t!]
\begin{center}
\includegraphics[width=0.325\textwidth]{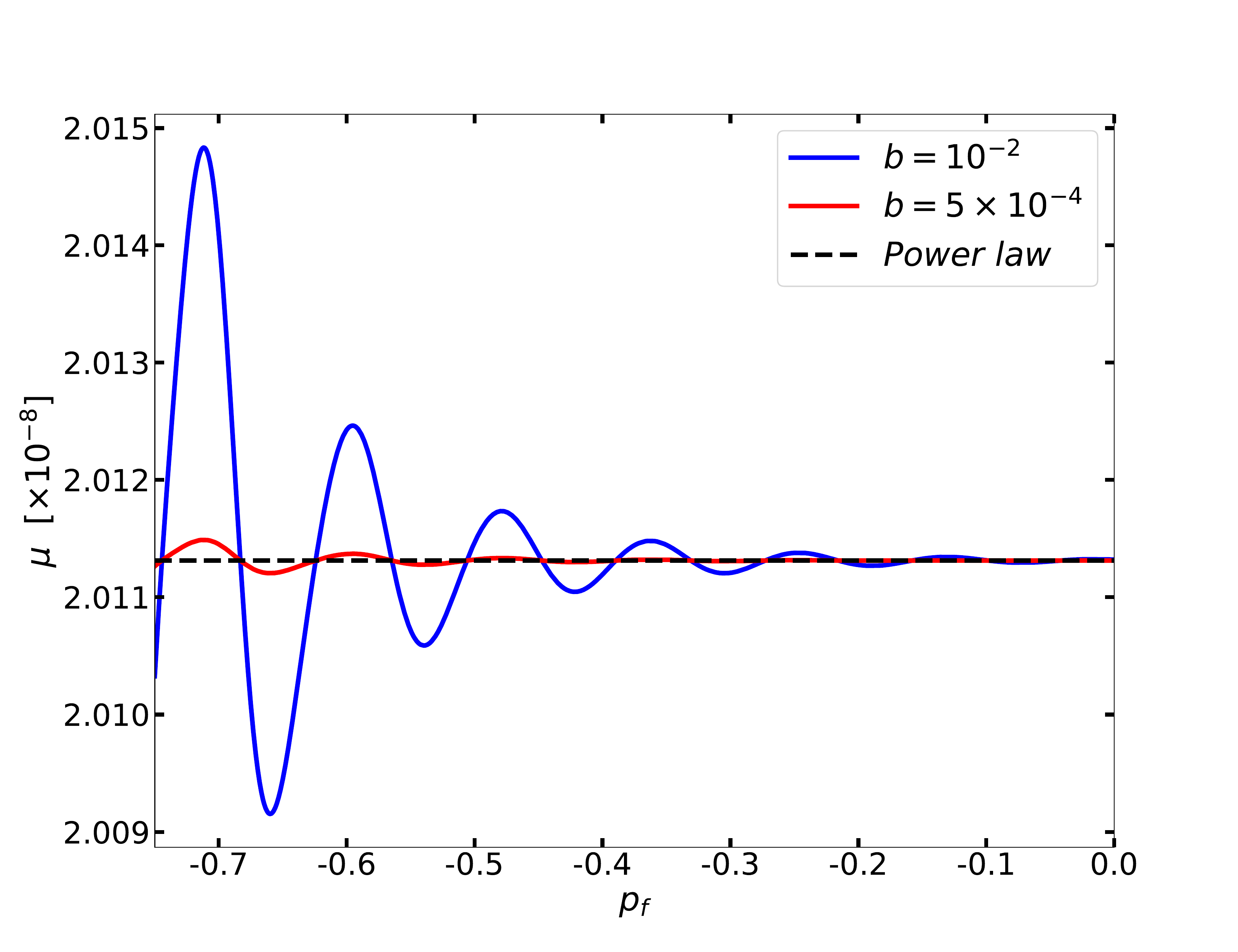}
\includegraphics[width=0.325\textwidth]{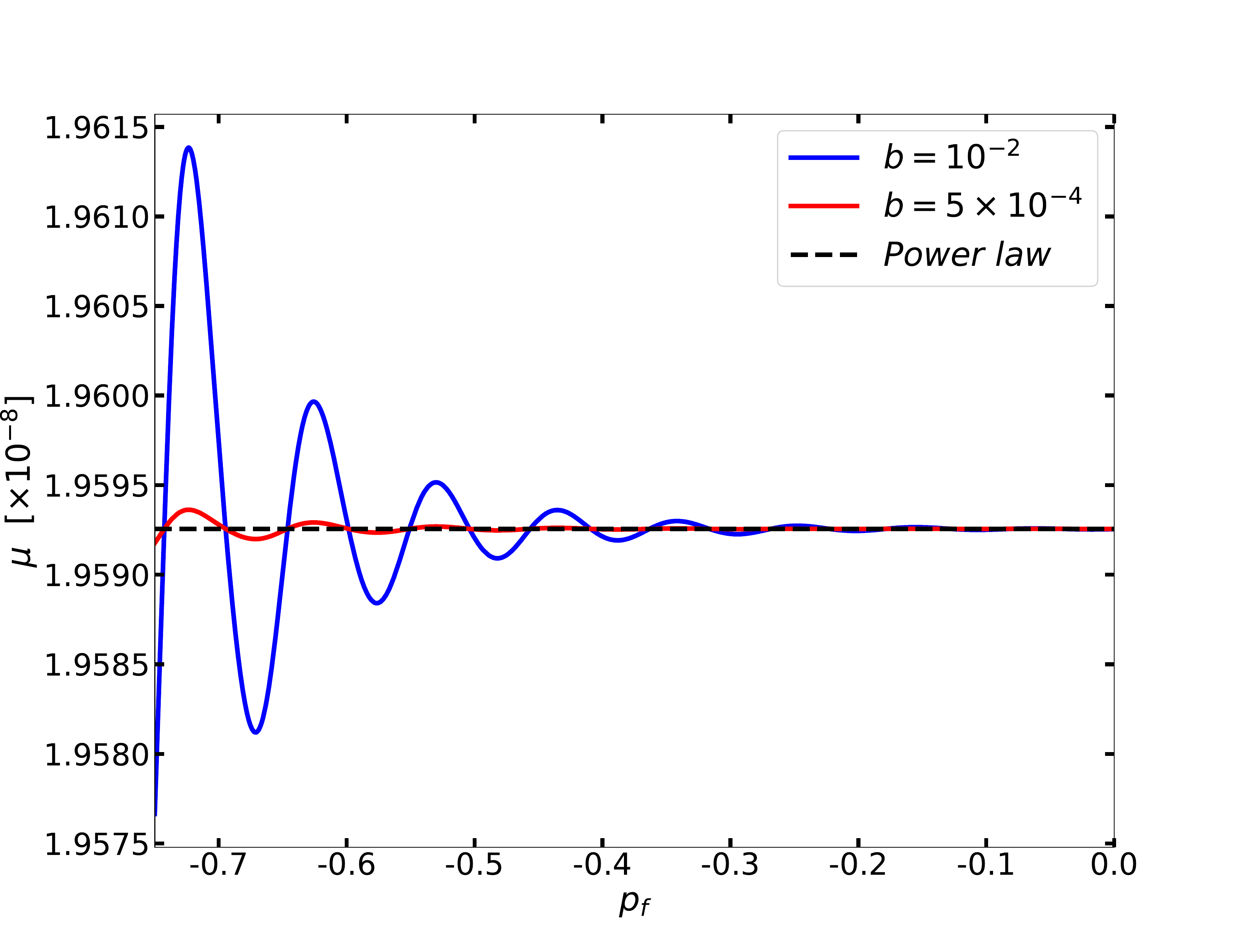}
\includegraphics[width=0.325\textwidth]{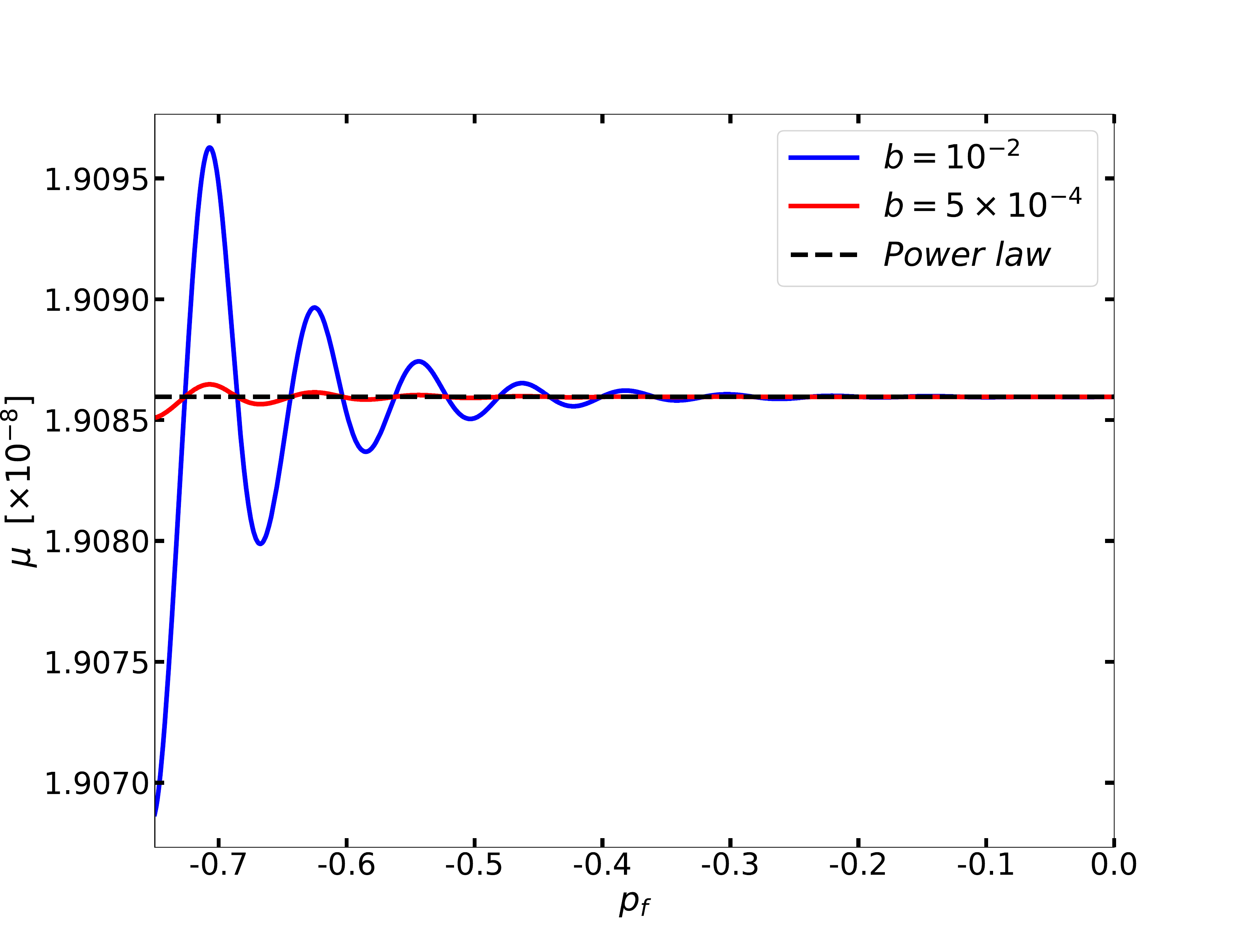}
\includegraphics[width=0.325\textwidth]{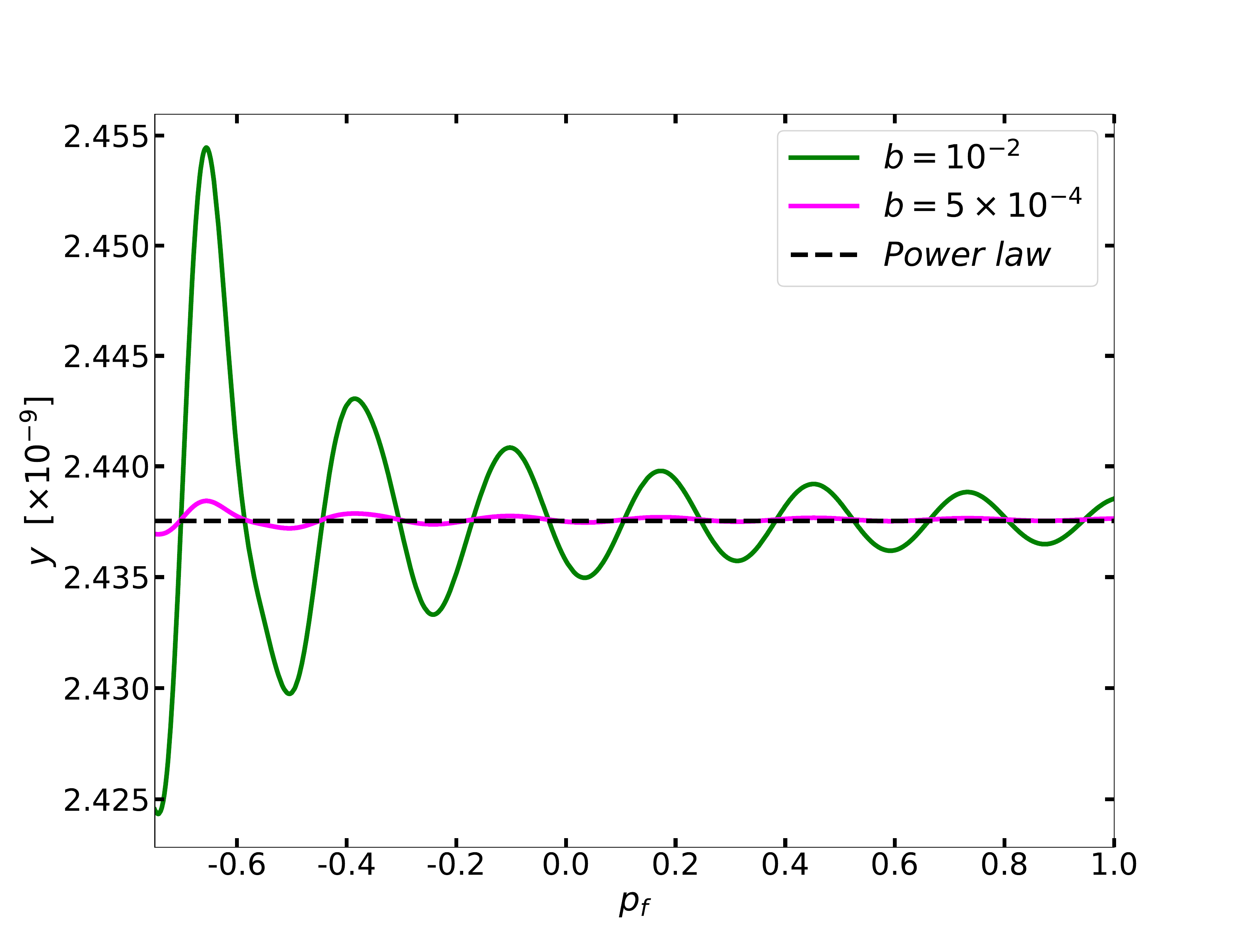}
\includegraphics[width=0.325\textwidth]{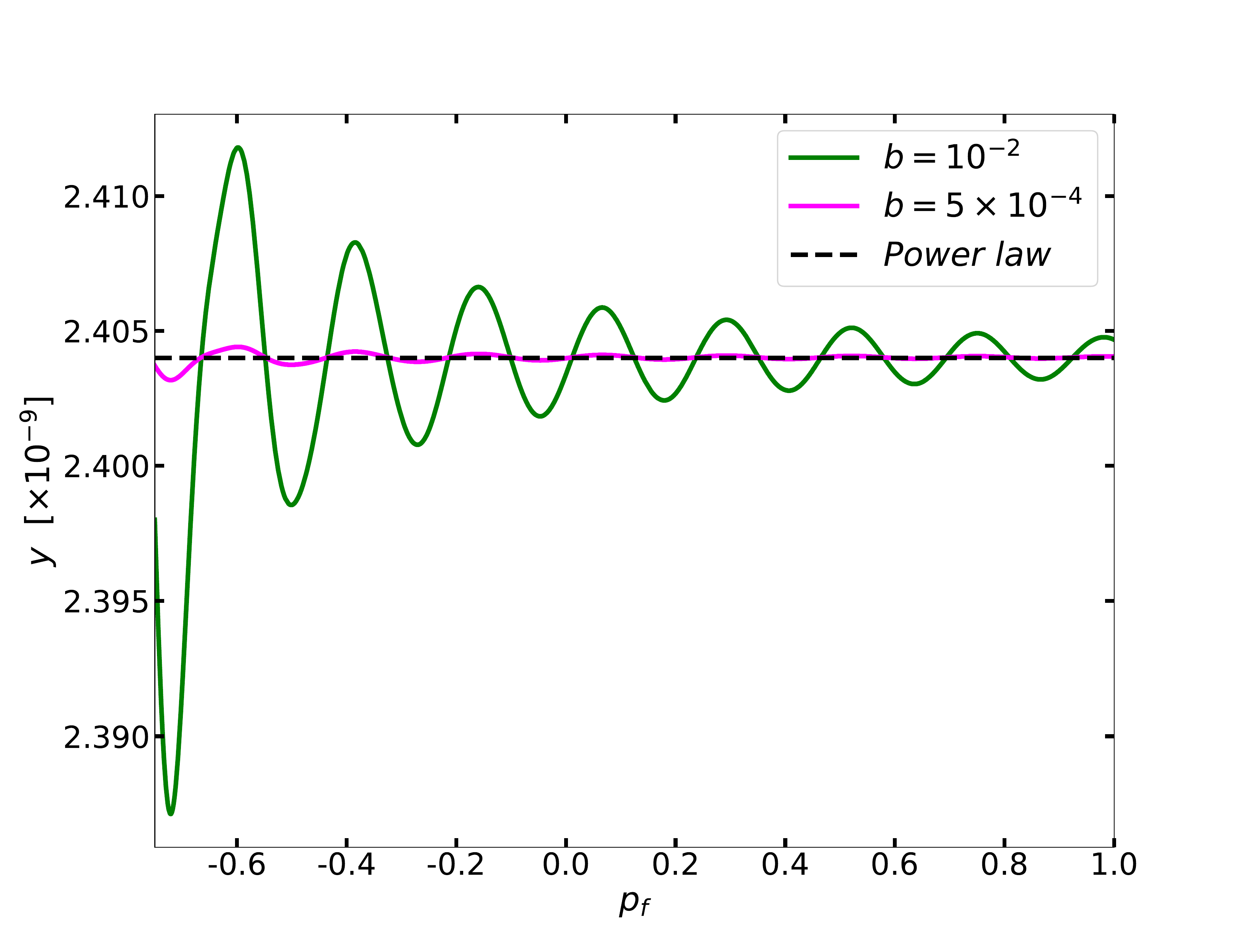}
\includegraphics[width=0.325\textwidth]{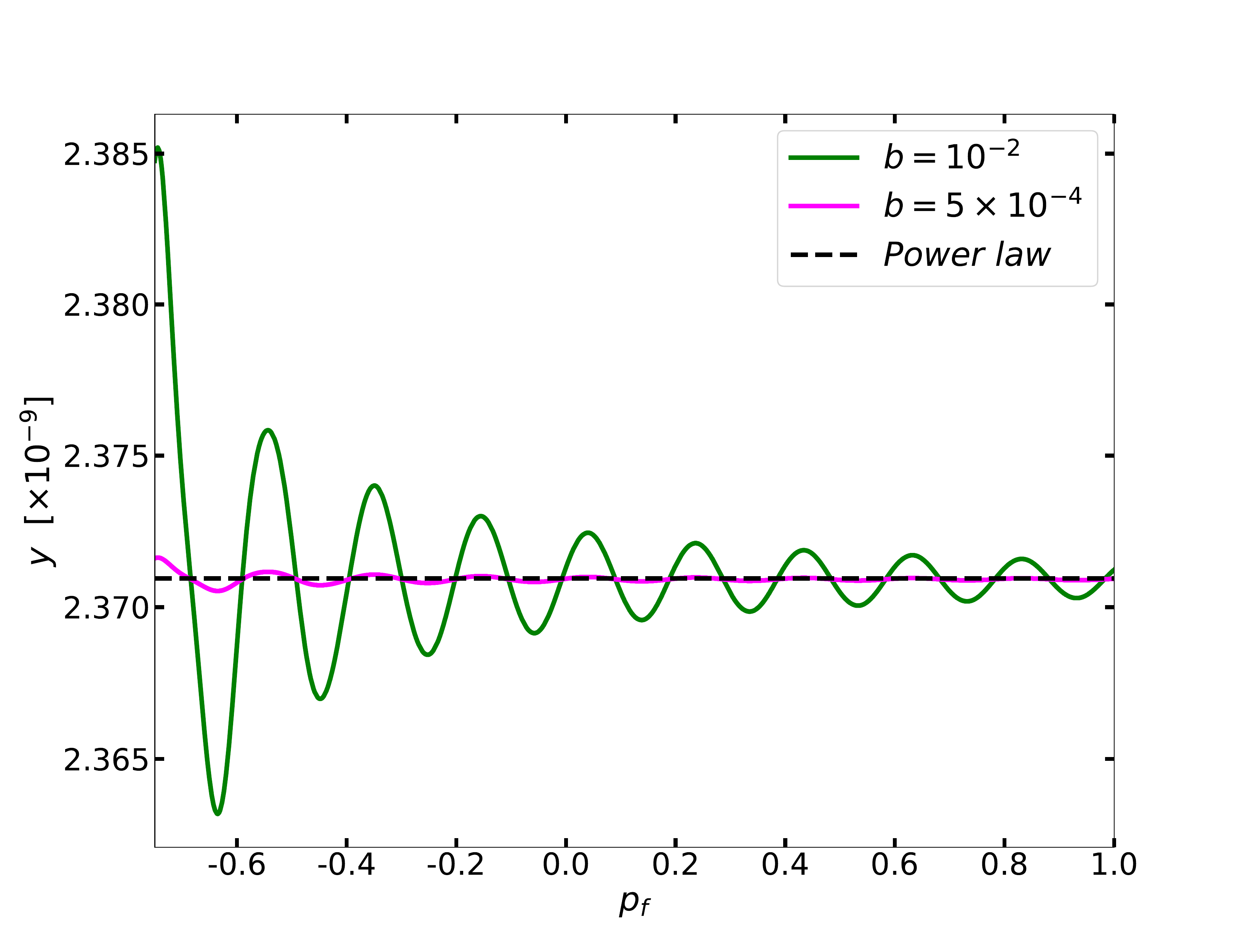}
\caption{$p_f$-dependence of $\mu$ (top) and $y$ (bottom) SD in models based on axion monodromy for different values of $p$: $p=\nicefrac23$ (left), $p=1$ (middle) and $p=\nicefrac43$ (right). A larger value of the modulation parameter $b$ yields larger SD, cf.~eq.~\eqref{eq:pkmndy1}. As $p_f$ grows, the oscillatory behavior of SD in axion-monodromy models tends to the values of a standard power-law potential with $b=0$. We assume here a fixed axion decay constant, $f_0=0.01$.}
\label{plot:axm3}
\end{center}
\end{figure}

Based on the results displayed in figure~\ref{plot:axm2}, in order to better appreciate the details of the SD and their dependence on $p_f$, we consider now a choice of the axion decay constant $f_0$ and the modulation parameter $b$ and vary $p_f$. We present our results in figure~\ref{plot:axm3} for $f_0=0.01$. The results show an interesting wave-damping behavior for the benchmark values of the monomial power $p=\nicefrac23$ (left panel), $1$ (center panel) and $p=\nicefrac43$ (right panel). For our selected value of $f_0$ we take the maximally allowed value $b=0.01$ (blue for $\mu$ and green for $y$ SD) and the minimally explored value $b=5\times10^{-4}$ (red for $\mu$ and magenta for $y$ SD). As expected from eq.~\eqref{eq:pkmndy1}, smaller values of $b$ lead to smaller SD. Note that the smaller $p_f$ is, the larger the SD become. The central value (horizontal dashed line) represents the magnitude of the SD for a standard power-law potential with $b=0$ for each value of $p$. In average, for $p_f=-0.7$ the SD are $0.2\%$ for $\mu$ and $0.7\%$ for $y$ larger than in the standard $b=0$ case.
The oscillatory behavior of SD in axion monodromy tends to the central value for larger positive values of the drifting power $p_f$. Recalling that this parameter encodes the possible interactions between the inflation axion and background moduli fields (or similar) of the full model if some dynamics among such fields is  left, variations of $p_f$ could happen and, hence the oscillatory behavior displayed in figure~\ref{plot:axm3} might be observable.

\begin{figure}[t!]
\begin{center}
\includegraphics[width=0.325\textwidth]{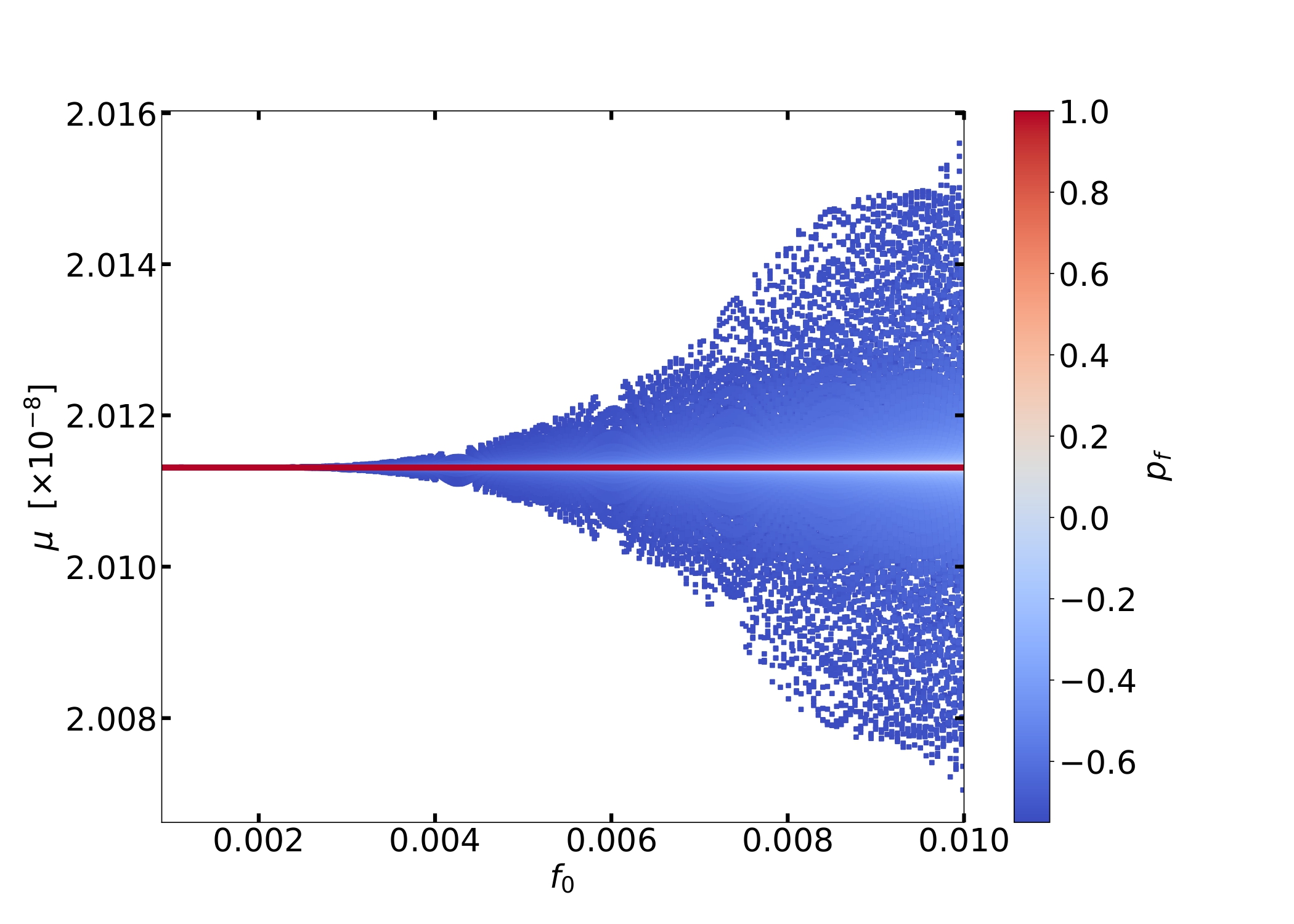}
\includegraphics[width=0.325\textwidth]{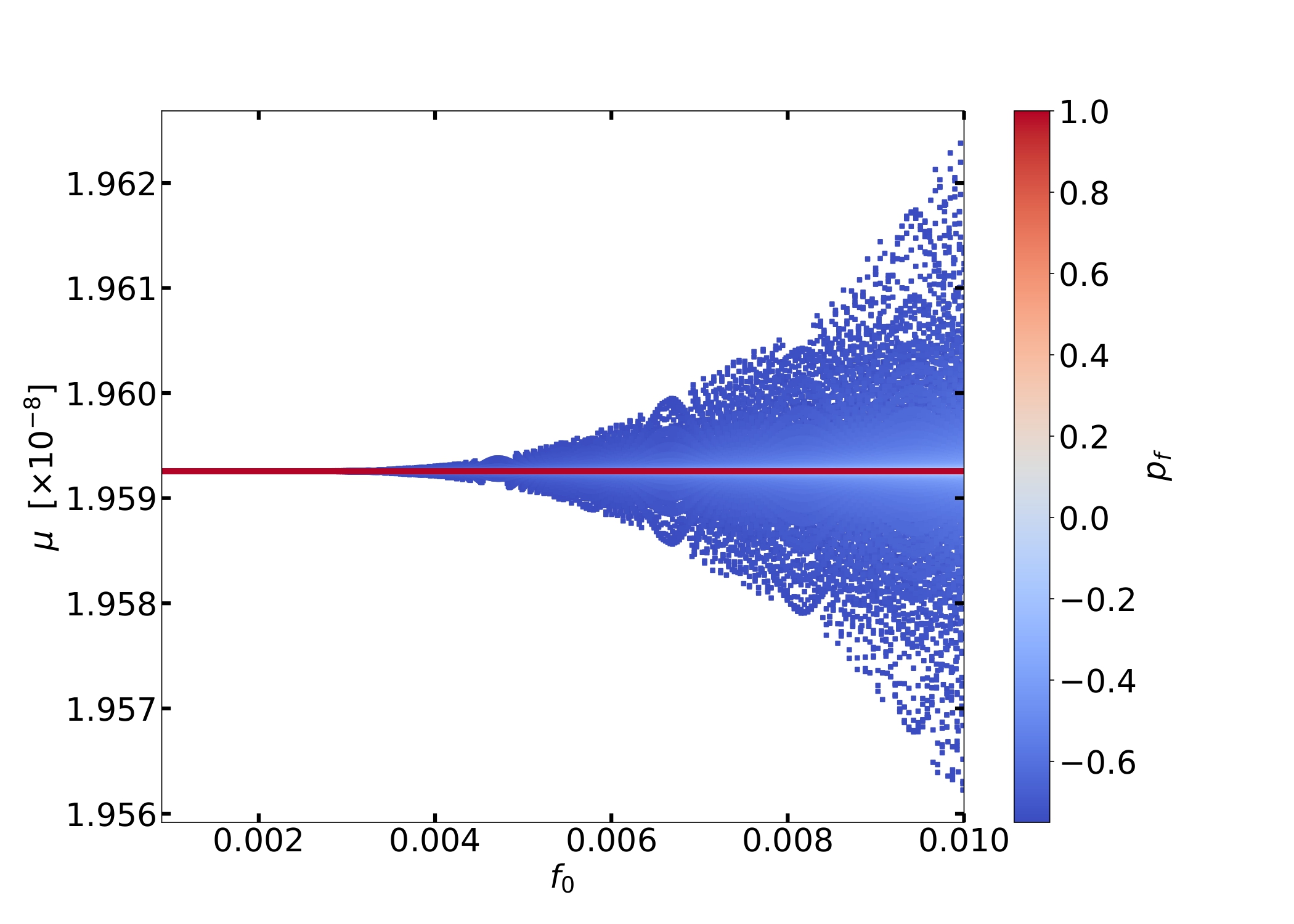}
\includegraphics[width=0.325\textwidth]{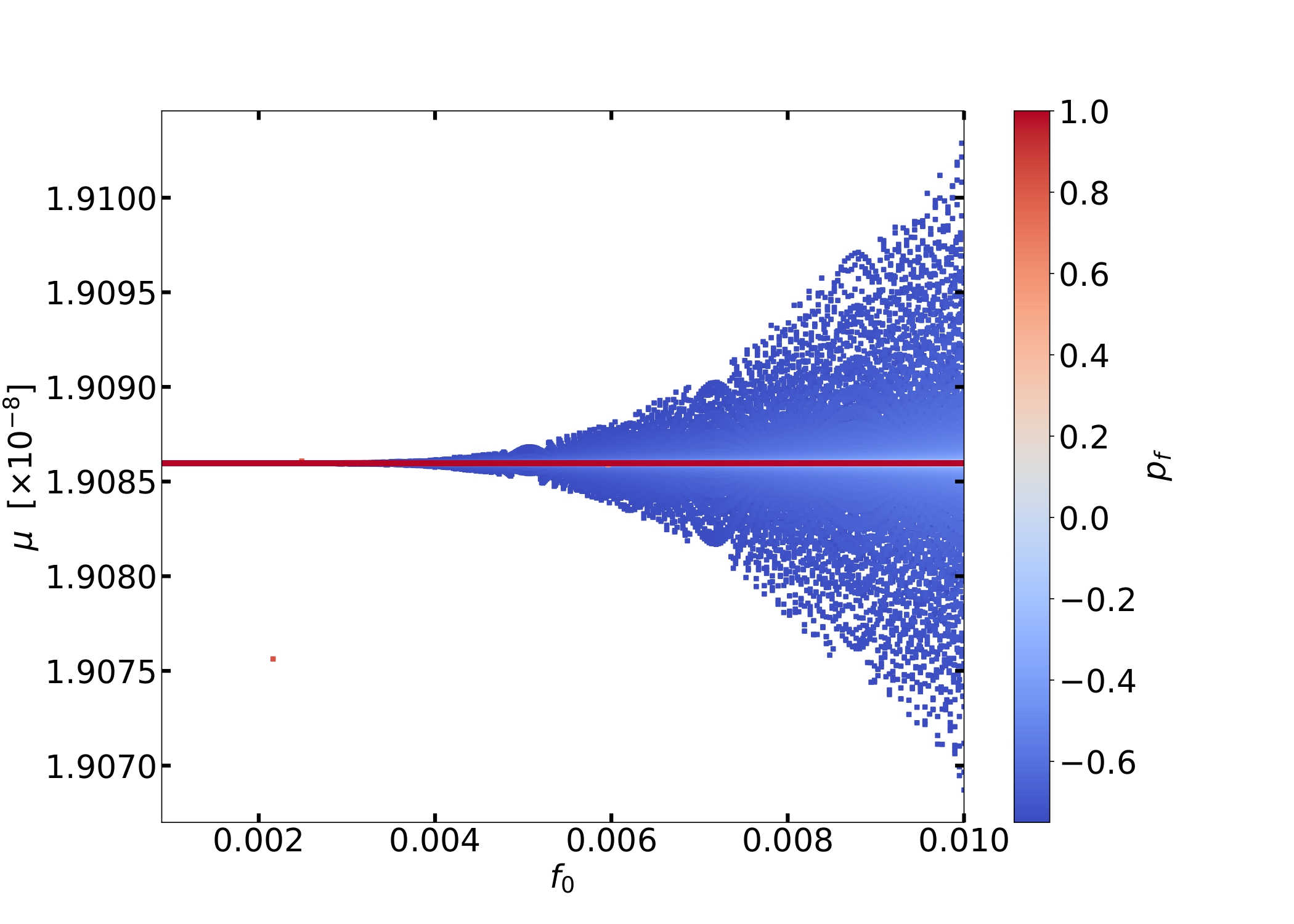}
\includegraphics[width=0.325\textwidth]{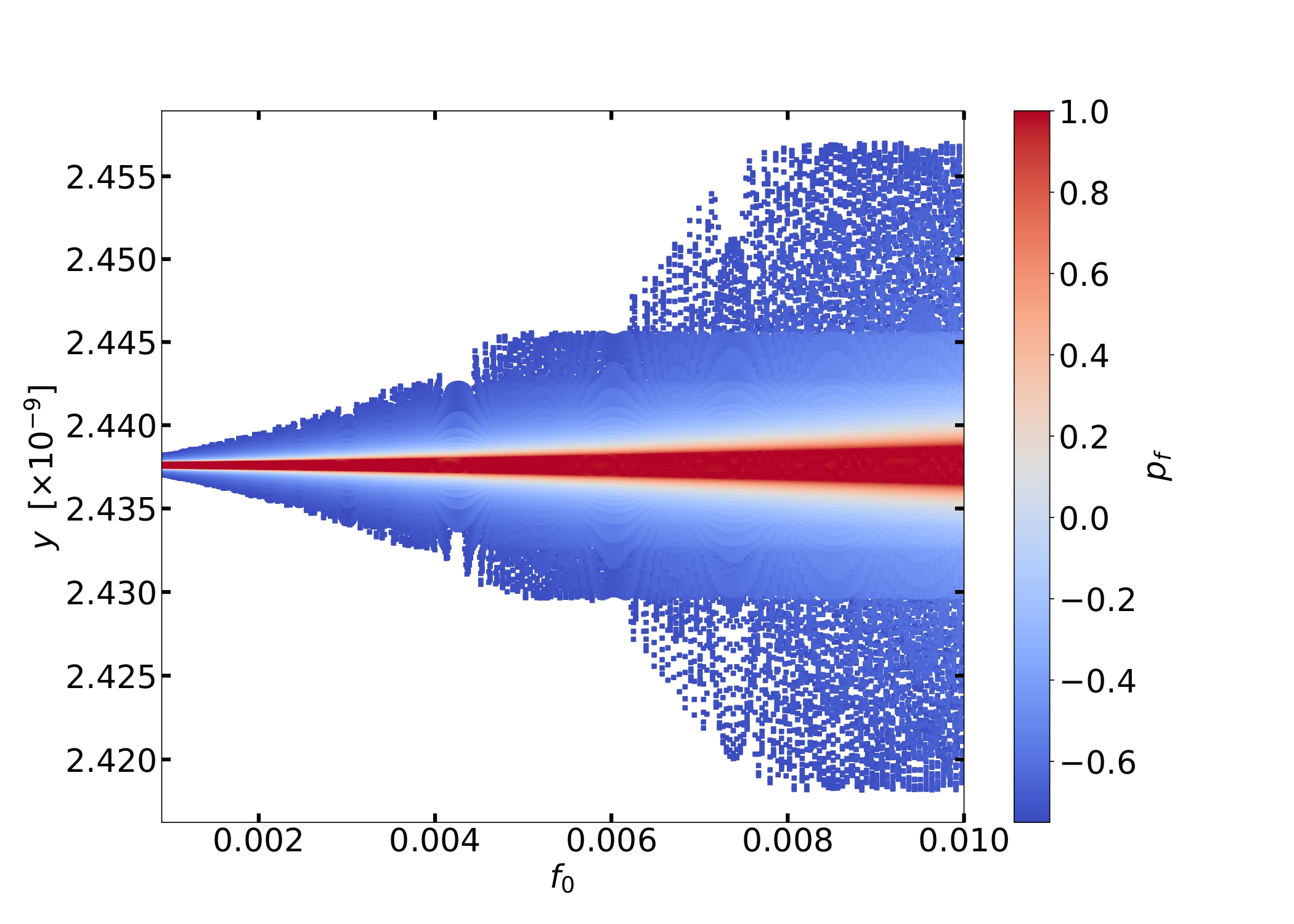}
\includegraphics[width=0.325\textwidth]{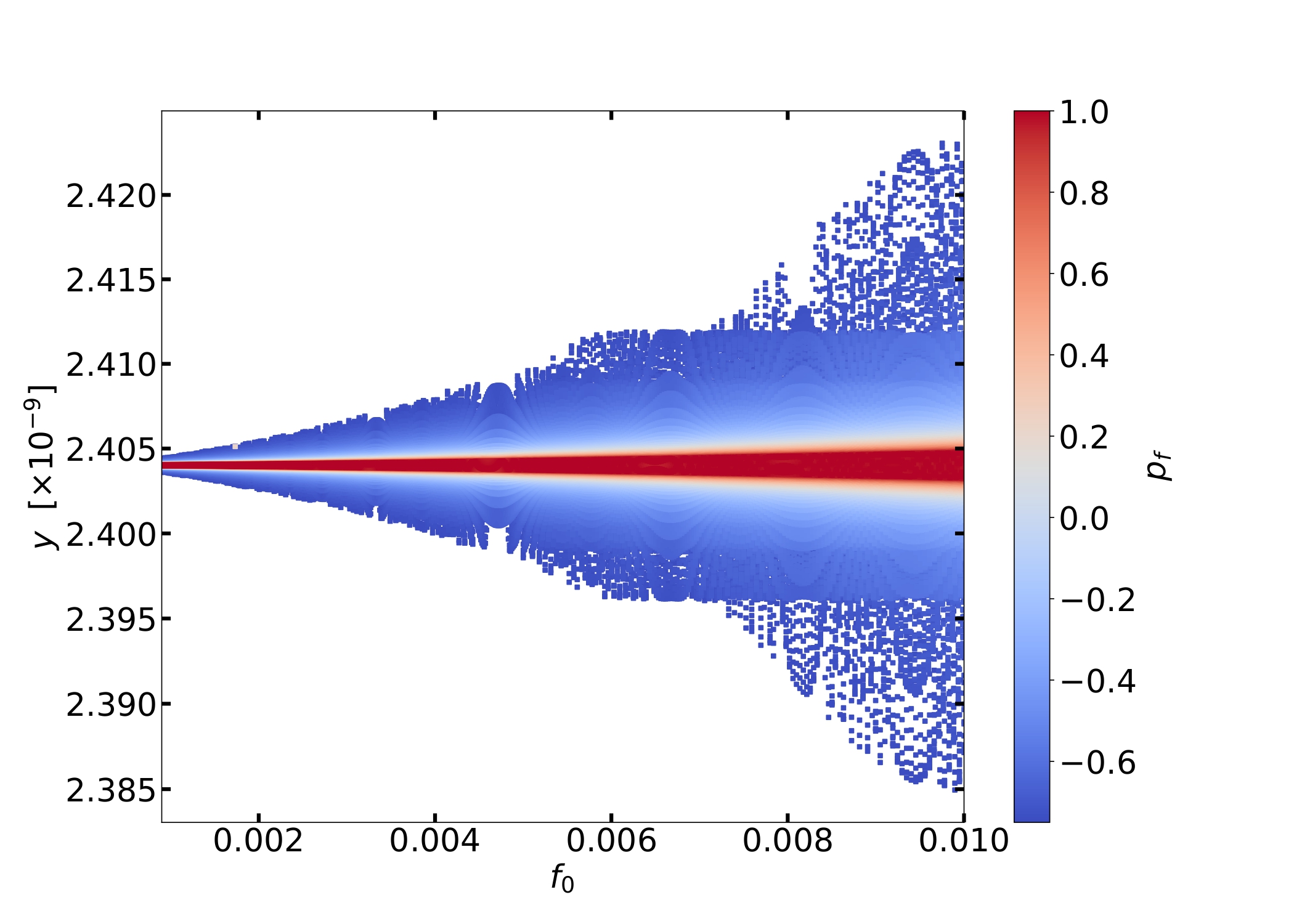}
\includegraphics[width=0.325\textwidth]{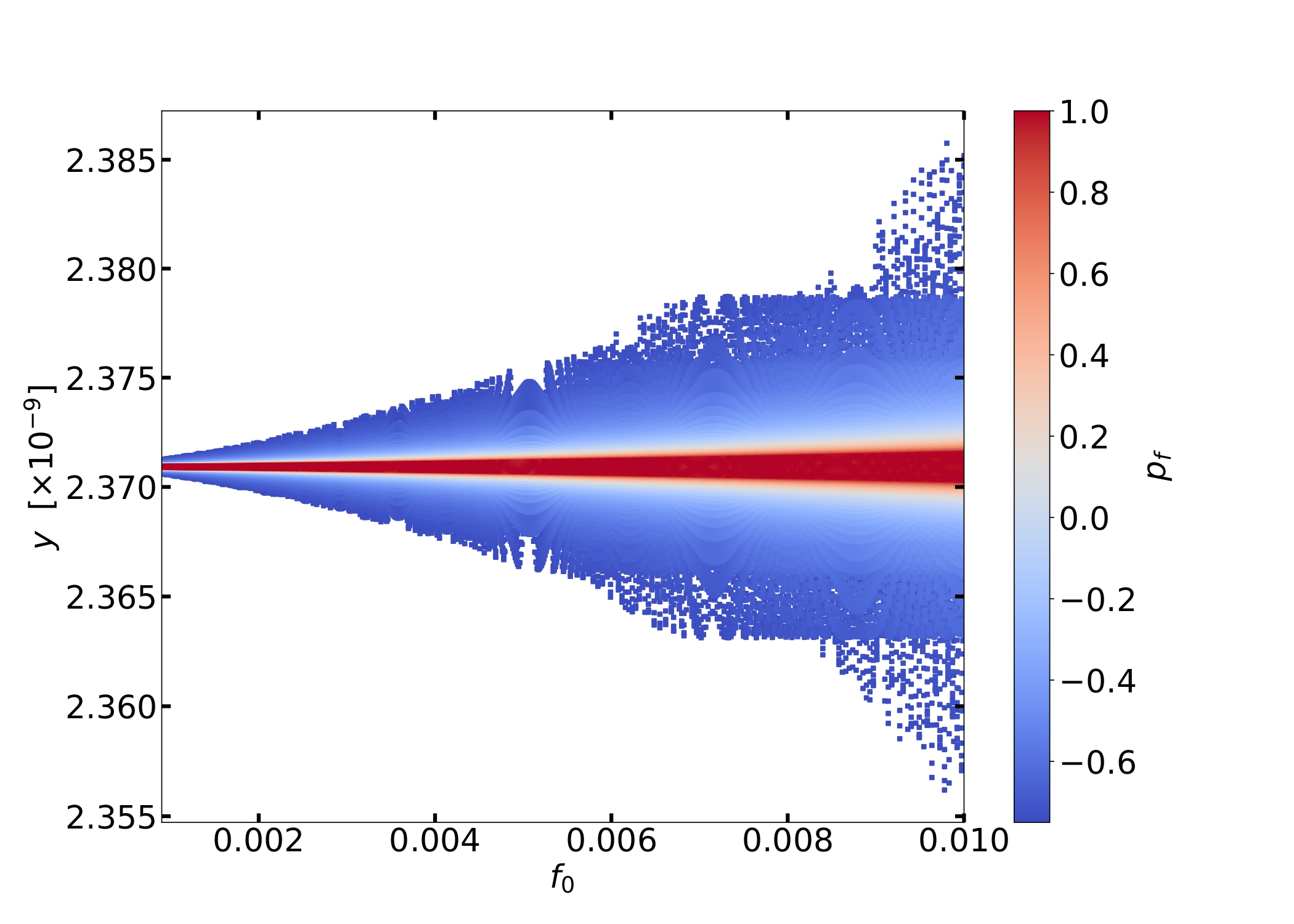}
\caption{Values of $\mu$ (top) and $y$ (bottom) SD as functions of the axion decay constant $f_0$ and the frequency drift $p_f$ in inflationary models based on axion monodromy. We have set the modulation parameter $b=0.01$ and display the results for $p=\nicefrac23$ (left), $p=1$ (middle) and
$p=\nicefrac43$ (right). For smaller values of $f_0$ and/or values of $p_f$ close to unity, the results from axion monodromy and standard power law with $V\sim\phi^p$ are very similar.}
\label{plot:axm4}
\end{center}
\end{figure}

Let us now include the variation of the axion decay constant $f_0$ in the scheme. We show in  figure~\ref{plot:axm4} all predicted values of $\mu$ and $y$ SD by axion-monodromy models. As before, we display on the top panels the values of the $\mu$ SD and in the bottom the values for $y$. Further, from left to right, we show the results for $p=\nicefrac23,1$ and $\nicefrac43$. The heatmap allows us to appreciate the effect on the SD of the variation of $p_f$. An interesting observation is that choosing $p_f\approx1$ or small values of $f_0$ lead to SD values that coincide with the central value obtained from a standard power-law potential (with $b=0$). This implies a conservative theoretical bound for detection of SD due to axion monodromy, distinguishable from power law, at around $f_0 \gtrsim 4\times10^{-3}$ and $p_f \lesssim 0.2$. Note that there are deviations from the central value of $y$ SD for smaller $f_0$; however, $y$ SD are in general an order of magnitude smaller than $\mu$ SD and hence leave a much weaker imprint in the SD signals.

Aiming at the detectability of our scenario, we focus on the parameter values that produce the largest SD signals and are compatible with Planck data. By inspecting figures~\ref{plot:axm2}, \ref{plot:axm3} and~\ref{plot:axm4}, we realize that $b=0.01=f_0$ and $p_f\approx-0.7$ render the most sizable signals. For each benchmark $p$ value, we show the resulting maximal values of $\mu$ and $y$ SD in table~\ref{table:sdam}. To establish a comparison with the standard power-law case ($b=0$), we also present in the table the SD values for all $p$ values. We find small but important enhancements of about $0.2-0.7$\% in axion monodromy over the standard power-law scenario.

\begin{table}[t!]
    \centering
    \begin{tabular}{c|cc|cc}
        & \multicolumn{2}{c|}{axion-monodromy SD} &
          \multicolumn{2}{c}{power-law SD}\\
    $p$ & $10^8\max(\mu)$ &   $10^9\max(y)$  & $10^8\mu$ & $10^9y$  \\
    \hline
     $\nicefrac23$  & $2.0148$  & $2.4545$ & $2.0113$ & $2.4376$ \\
     $1$    & $1.9614$ & $2.4118$   & $1.9592$ & $2.4039$ \\
     $\nicefrac43$   & $1.9096$   & $2.3852$ & $1.9085$& $2.3708$
    \end{tabular}
    \caption{Comparison between the maximal values of SD in axion monodromy (with $b=0.01$) and power-law inflationary models ($b=0$). The maximal values of $\mu$ SD occur at $p_f=\{-0.71, -0.72, -0.71\}$ and of $y$ at $p_f=\{-0.66, -0.59, -0.74\}$, for $p\in\{\nicefrac23,1,\nicefrac43\}$, respectively. In all cases, we have taken $f_0=0.01$ and $b=0.01$.}
    \label{table:sdam}
\end{table}

\begin{figure*}[t!]
\begin{center}
\includegraphics[width=0.45\textwidth]{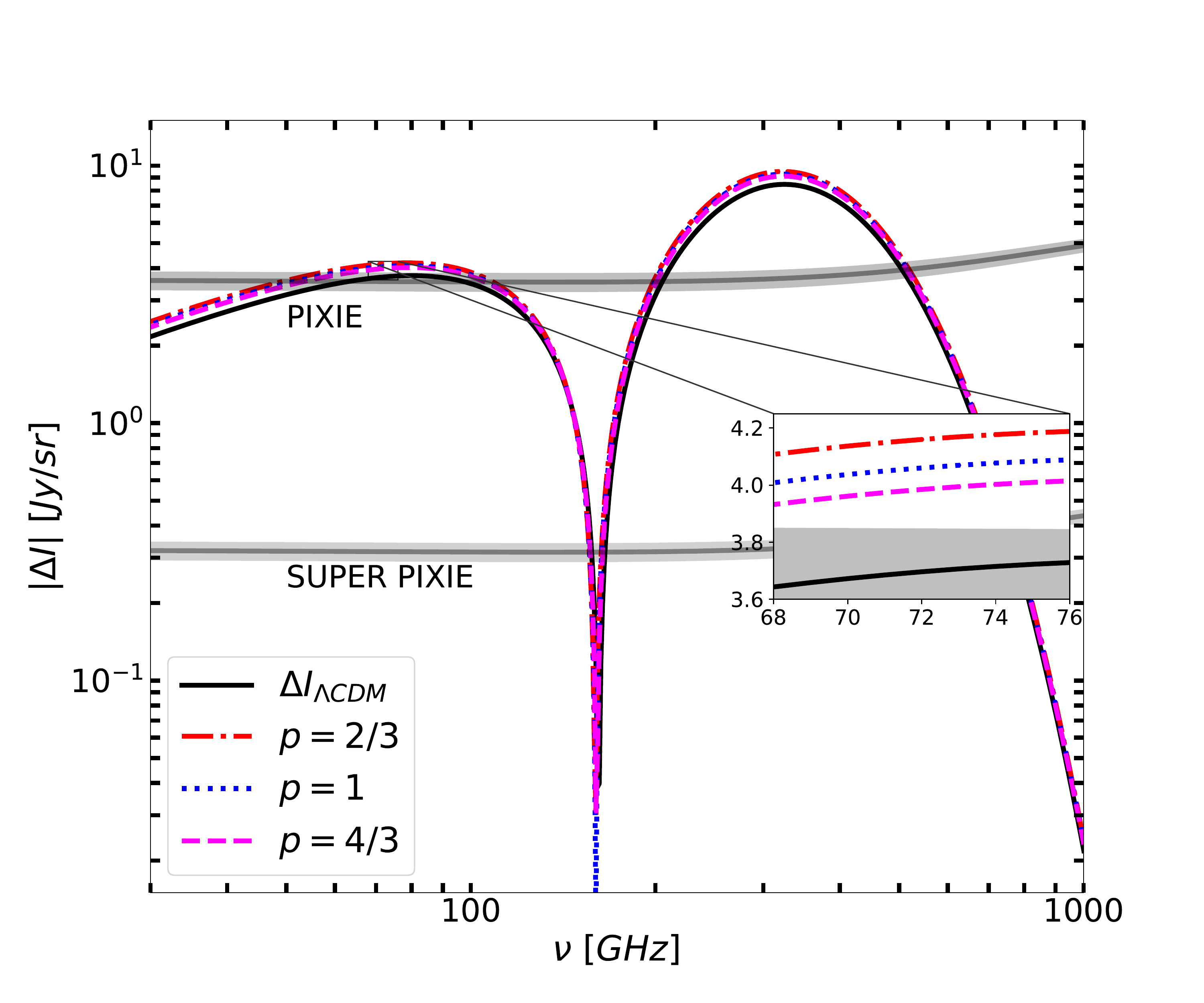}
\includegraphics[width=0.45\textwidth]{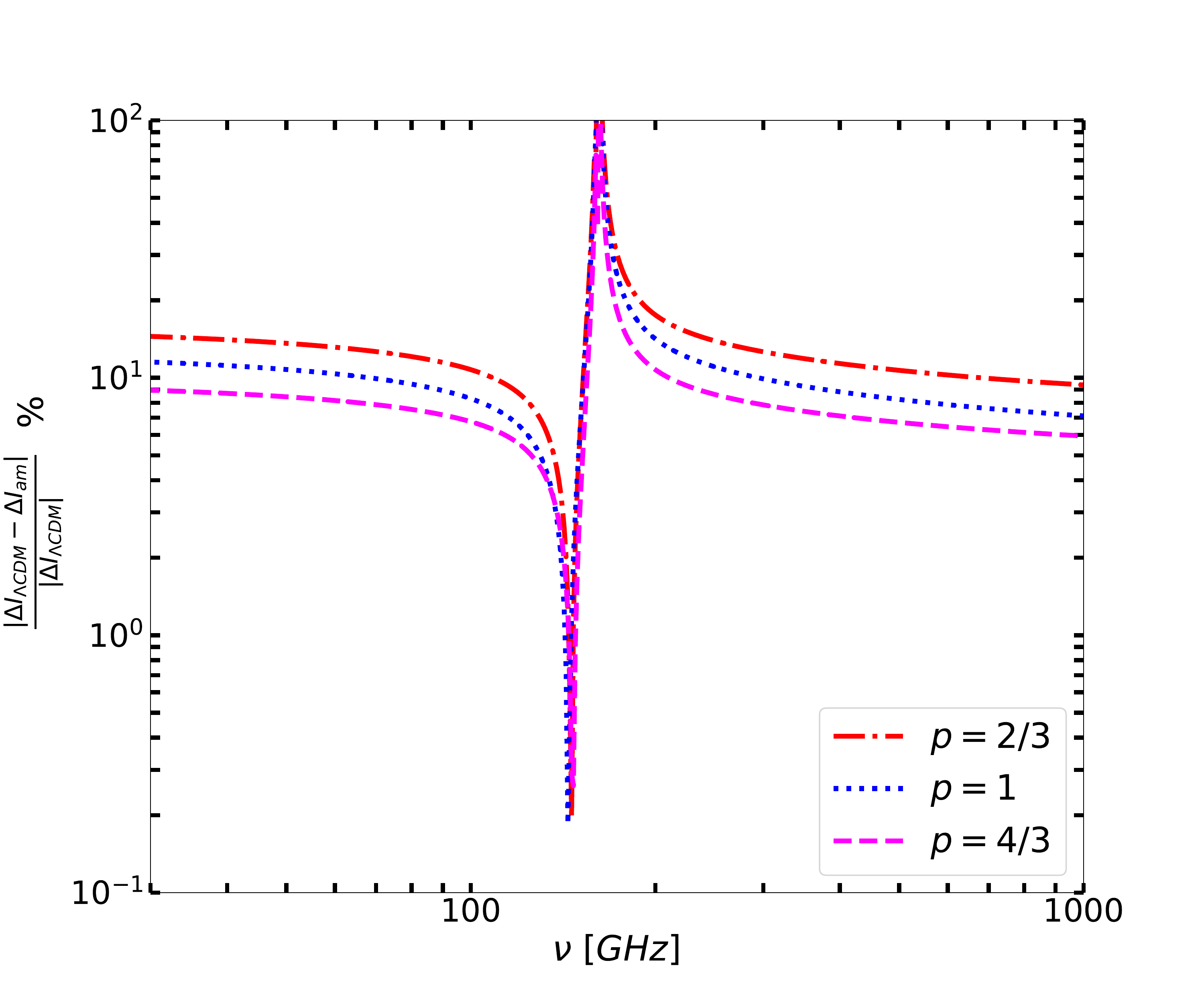}
\caption{Predictions on the contribution to the distortion $\Delta I$ of the photon intensity~\eqref{eq:sd1} arising from $\mu$ and $y$ SD in axion monodromy inflationary models, contrasted against the $\Lambda$CDM prediction. To explore the maximal size of the SD contributions from axion monodromy complying with Planck bounds, we take $p_f=-0.7$, $b=0.01$ and $f_0=0.01$. In the left, we display the three curves described for the benchmark values of monomial power, $p=\nicefrac23$ (red dash-dotted curve), $p=1$ (blue dotted) and $p=\nicefrac43$ (magenta dashed), and the $\Lambda$CDM prediction (black continuous curve). We further show the sensitivity of future PIXIE and Super PIXIE missions (curves adapted from~\cite[fig.~9]{Chluba:2019nxa}). In the right, we evaluate the difference between the axion-monodromy $\Delta I_\mathrm{am}$ and $\Lambda$CDM $\Delta I_{\Lambda\mathrm{CDM}}$ predicted intensities for the different benchmark $p$ values. For frequencies outside the range $100\,\mathrm{GHz}\lesssim\nu\lesssim250\,\mathrm{GHz}$ the difference w.r.t.\ the fiducial signal is about 10\%.}
\label{plot:axm5}
\end{center}
\end{figure*}

To conclude this section, we now address the question of whether the SD predictions associated with inflationary models based on axion monodromy can be falsified by future observational data. With this goal in mind, first we compute the $\mu$ and $y$ contributions to the distortions of the photon intensity spectrum $\Delta I (\nu)$, see eq.~\eqref{eq:sd1}, and compare the results to the sensitivity of the PIXIE experiment and its enhanced version Super-PIXIE. In figure~\ref{plot:axm5} we plot in the left the intensity in units of $\mathrm{Jy}/\mathrm{sr}=10^{-26}\, \mathrm{W}\,\mathrm{m}^{-2}\mathrm{Hz}^{-1}\mathrm{sr}^{-1}$ for axion monodromy with $p=\nicefrac23$ (red dash-dotted curve), $p=1$ (blue dotted), $p=\nicefrac43$ (magenta dashed curve), and for the $\Lambda$CDM model (black continuous curve). The latter is obtained from inserting the observed values of $n_s=0.96605\pm0.0042$~\cite[Table 1]{Planck:2018vyg} and $\mathcal A_s$ in the standard primordial power spectrum, eq.~\eqref{eq:gralPowerSspectrum}, with $\alpha_s=0$ and then computing the SD as in eq.~\eqref{eq:sd5}. We observe that in the range $55\lesssim\nu\lesssim110$\,GHz axion monodromy would leave an observable SD signal whereas SD from $\Lambda$CDM would not be detectable.
Moreover, to quantify the difference between SD from axion monodromy (am) and from $\Lambda$CDM, we compute $|\Delta I_{\Lambda\mathrm{CDM}}-\Delta I_\mathrm{am}|$ and express this difference on the right-hand side of figure~\ref{plot:axm5} as a percentage of the $\Lambda$CDM result. We see that they can differ by up to about $10$\% in the physically relevant region, $\nu\lesssim100$\,GHz and $\nu\gtrsim250$\,GHz. Interestingly, the greatest discrepancy, though marginal, is realized for axion monodromy with $p=\nicefrac23$.

A second important observation we need to study the falsifiability of axion monodromy is the experimental error of future measurements of SD. As mentioned earlier, the expected standard error of Super-PIXIE is given by $\sigma(\mu)\simeq7.7\times10^{-9}$ and $\sigma(y)\simeq1.6\times10^{-9}$~\cite{Chluba:2019nxa, Delabrouille:2019thj}. Unfortunately, our comparison in table~\ref{table:sdam} between power-law and axion monodromy indicates that we need $\sigma(\mu),\sigma(y)\sim10^{-11}$ to distinguish between those two scenarios. This can be achieved by the proposed configurations of PIXIE that shall enhance its sensitivity by a factor of 100~\cite{Fu:2020wkq}.

%%%%%%%%%%%%%%%%%%%%%%%%%%%%
%%%%%%%%% ns and r %%%%%%%%%
\section[Planck constraints on ns and r]{\boldmath Planck constraints on $n_s$ and $r$ \unboldmath}
\label{sec:ns-r}
%%%%%%%%%%%%%%%%%%%%%%%%%%%%

\begin{table}[t!]
    \centering
    \begin{tabular}{c|c|ccc|cc}
    $p$     & $\gamma_0$ & $\phi_\star$ & $\phi_\mathrm{end}$ & $10^4 \lambda$ & $n_s$ & $r$\\
    \hline
    $\nicefrac23$ &  $2.00$ & $8.78$ & $0.67$ & $14.38$ & $0.965$ & $0.046$\\
    $1$ & $2.78$ &  $10.77$ & $0.99$ & $5.85$ & $0.965$ & $0.069$\\
    $\nicefrac43$ &  $4.54$ & $12.38$ & $0.01$ & $1.79$ & $0.965$ & $0.093$
    \end{tabular}
    \caption{Improved parameters of axion-monodromy inflation with fixed values of $b = f_0 = 0.01$, $p_f = -0.7$, $k_\star=0.05$\,Mpc$^{-1}$ and $N_\star=57.5$. Both the inflaton field $\phi$ and the scale $\lambda$ are given in units of Planck mass. By varying the value of the phase $\gamma_0$ (in radians) in eq.~\eqref{eq:potential}, it is possible to successfully fit the spectral tilt $n_s$ while keeping the tensor-to-scalar ratio $r$ within the $3\sigma$ C.L. observed region, cf. table~\ref{tab:ModelParameters}.}
    \label{tab:ModelParameters2}
\end{table}

Since axion monodromy is known to yield large tensor modes, current bounds on the spectral tilt $n_s$ and the tensor-to-scalar ratio $r$ are additional observables that can be used to falsify inflationary models based on axion monodromy. In this section, we briefly revise the status of the model on this topic.

\begin{figure*}[t!]
\begin{center}
\includegraphics[width=0.495\textwidth]{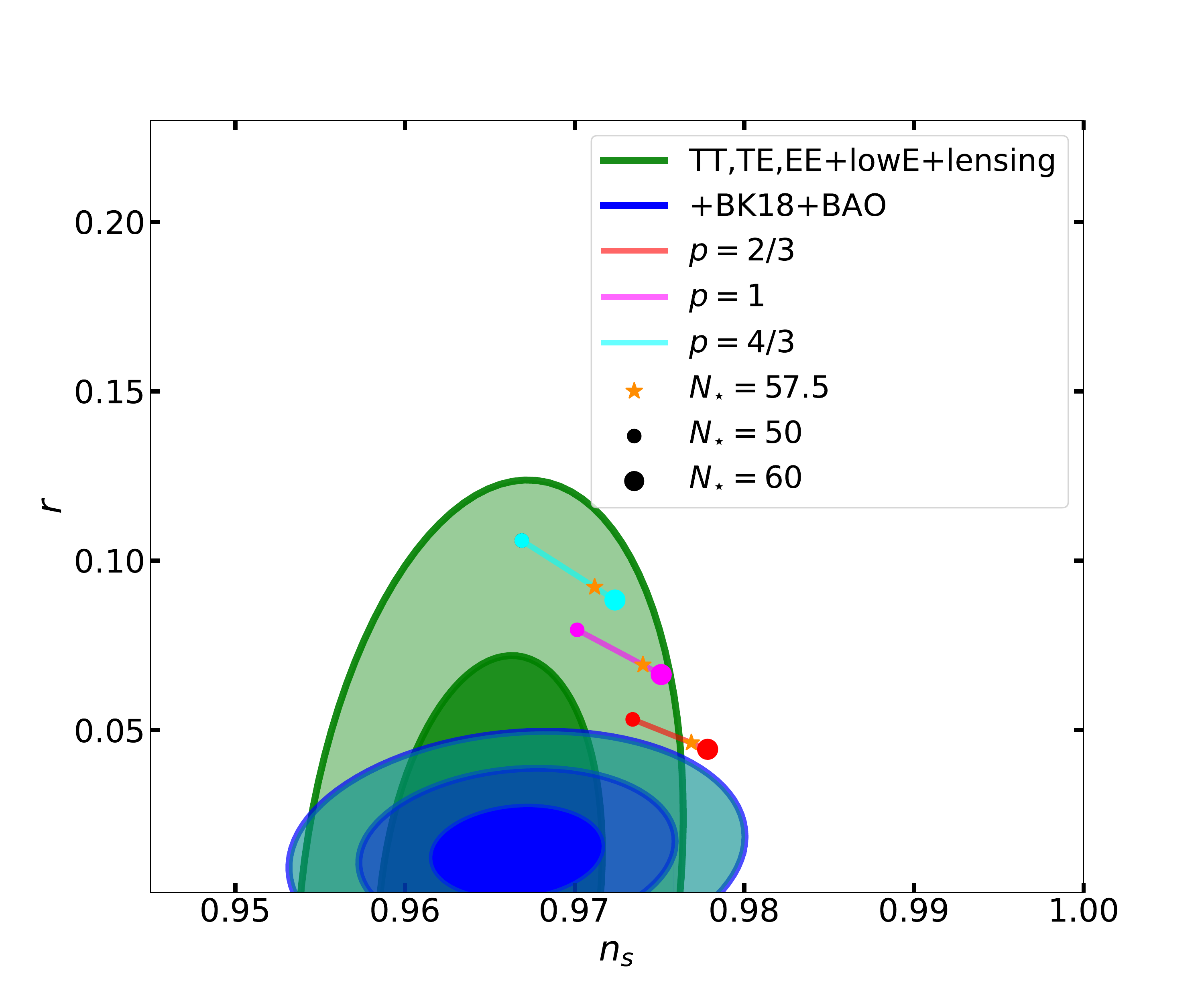}
\includegraphics[width=0.495\textwidth]{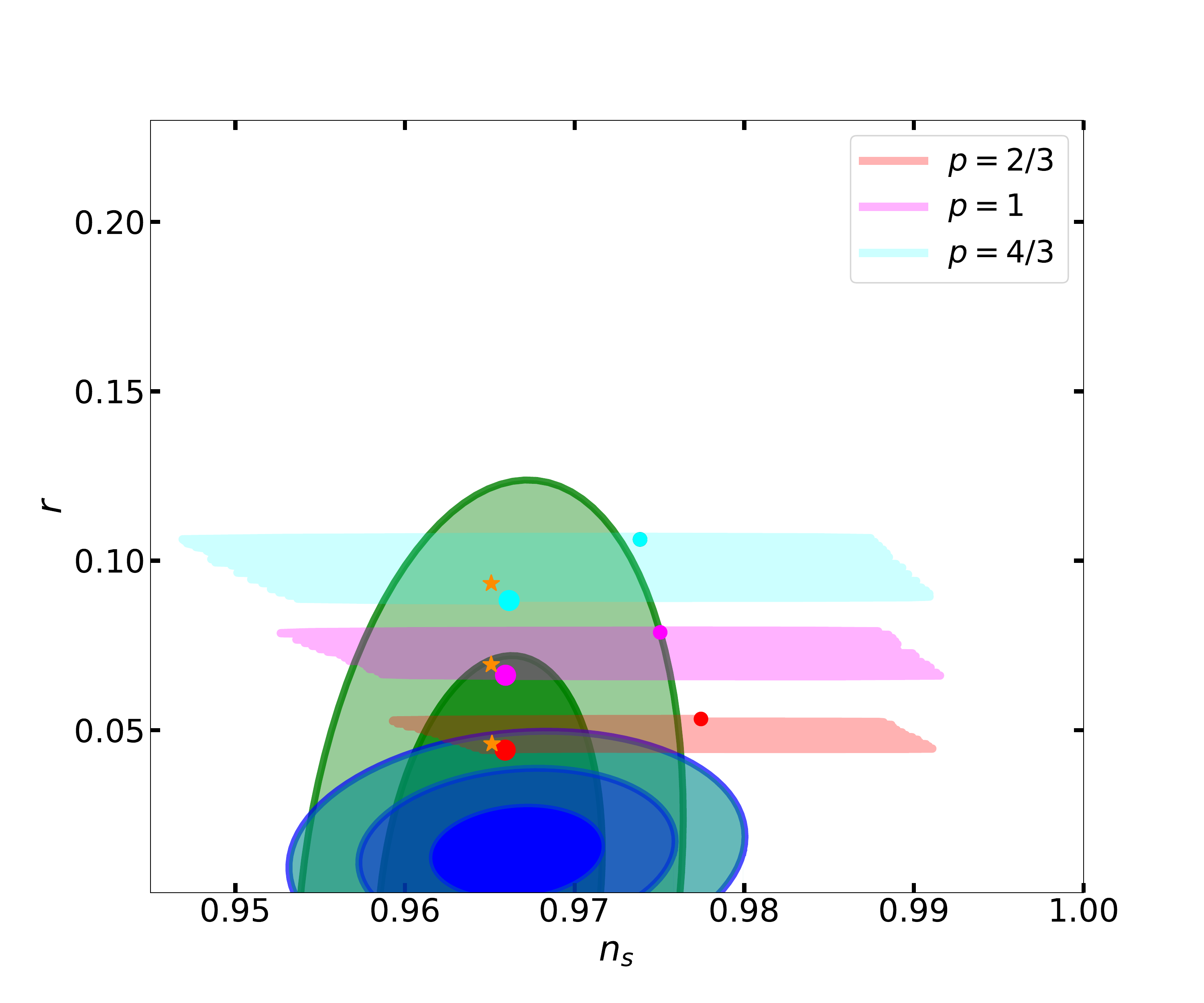}
\caption{Predictions for $n_s$ and $r$ from $\phi^p$ (left) and axion monodromy inflation (right) with $50\leq N_\star\leq60$ e-folds contrasted against the observational constraints from the CMB Planck data in combination with baryon acoustic oscillation (BAO), CMB lensing data and BICEP/Keck data. The (higher) green contours correspond to the $1\sigma$ and  $2\sigma$ C.L.\ regions based on Planck's CMB data (TT,TE,EE$+$lowE) and lensing~\cite{Planck:2018vyg}. The (lower) blue contours depict the $1\sigma$, $2\sigma$ and  $3\sigma$ C.L.\ regions resulting from combining the previous data with BICEP/Keck 18 (BK18) and BAO data~\cite{BICEP:2021xfz}. For axion monodromy (right), we used the values of  $\gamma_0=2.00$ ($p=\nicefrac23$), $2.78$ ($p=1$) and $4.54$ ($p=\nicefrac43$) in radians, which produce the best fit for $n_s$ at $N_\star=57.5$. Labels on the right panel are the as in the left one. }
\label{plot:axm0}
\end{center}
\end{figure*}

The latest Planck's best-fit value for the spectral tilt is $n_s=0.96605\pm0.0042$~\cite[Table 1]{Planck:2018vyg}, while the upper bound on the tensor-to-scalar ratio is about $r<0.123$ at 2$\sigma$ based only on Planck's TT,TE,EE$+$lowE$+$lensing data~\cite{Planck:2018vyg} (depicted by the green contours of figure~\ref{plot:axm0}), and $r<0.048$ at 3$\sigma$ C.L. based on the latest combination of Planck's data together with BICEP/Keck (BK18) and BAO data~\cite{BICEP:2021xfz} (depicted by the blue contours in figure~\ref{plot:axm0}).

Disregarding the periodic modulation of axion monodromy, from eq.~\eqref{eq:nsandr} we find that the resulting single-field monomial potential yields
\begin{equation}
n_s\approx1-(p+2)/2N_0\qquad\text{ and }\qquad
r\approx4p/N_0\,.
\end{equation} 
In this scenario, we note some (known) tension between the prediction of a model based on $V\sim \phi^p$ and the observations. The values of $n_s$ and $r$ for our benchmark values of $N_\star$ and $p\in\{\nicefrac23,1,\nicefrac43\}$ are presented in table~\ref{tab:ModelParameters}. In the left plot of figure~\ref{plot:axm0} we explore these results for other admissible values of e-folds\footnote{Note that various values of $N_\star$ can be associated with $k_\star=0.05$\,Mpc$^{-1}$ because $N_\star$ depends on many other (undetermined) parameters, see e.g.~\cite[eq.\ (3.11)]{Stein:2021uge}.} $N_\star$. The dumbbells in different colors illustrate the values predicted by such simplified model with three different values of $p$. The small (large) bullet corresponds to $N_\star=50$ ($N_\star=60$) e-folds and the star denotes our (arbitrarily chosen) benchmark value $N_\star=57.5$, which is frequently used in the literature.  In this approximated model, $n_s$ is found within the 3$\sigma$ region of the combined fit of Planck and BK18. However, although $r$ is within the $2\sigma$ region of Planck's data, it lies beyond the $3\sigma$ C.L. region of the latest combined data.

So far, we have set $b=0$ and hence ignored the oscillatory modulation of axion monodromy. Setting the small modulation parameter $b\neq0$ introduces important changes on the predictions for $n_s$ and $r$. First, $\phi_\star$, $\phi_{\rm end}$ and $N_\star$ depend on $b,f_0,\gamma_0$ and $p_f$ besides $p$ and can only be computed numerically using an iterative approach. For fixed values of $p,b,f_0,\gamma_0$ and $p_f$, the value of $\phi_\star$ has an oscillatory behavior. Consequently, also the values of $n_s$ and $r$ oscillate. The spectral tilt oscillates in a wide range of values while $N_\star$ varies a little, depending on $p$ and the angle $\gamma_0$ appearing in the potential~\eqref{eq:potential}. The tensor-to-scalar ratio, on the other hand, oscillates minimally, such that its value resembles the standard power-law result, which only depends on $p$ and $N_\star$. Interestingly, these properties pull $n_s$ and $r$ closer to the observed values. In the right plot of figure~\ref{plot:axm0}, we display all different values of $n_s$ and $r$ for $50\leq N_{\star}\leq 60$ and our three benchmark choices of $p\in\{\nicefrac23,1,\nicefrac43\}$, assuming fixed values\footnote{The values of $b,f_0,p_f$ are chosen to maximize the spectral distortions of the model, as we saw in section~\ref{sec:sdam}.} of $b=f_0=0.01$ and $p_f=-0.7$. 
The phase $\gamma_0$ has been chosen independently for each $p$ with the goal of best fitting $n_s$ at $N_\star=57.5$ to the observed value; the stars in the plot correspond to the values of $n_s$ and $r$ that we obtain for $p=\nicefrac43$ (top), $1$ (middle) and $\nicefrac23$ (bottom). We conclude that $p\leq\nicefrac23$ and nontrivial phases $\gamma_0$ lead to axion monodromy models compatible with current Planck observations at $1\sigma$, and at $3\sigma$ for Planck and BICEP/Keck 18 combined data.

%%%%%%%%%%%%%%%%%%%%%%%%
%%% SECTION CONCLUSIONS %%%
\section{Conclusions}
\label{sec:conclusion}
%%%%%%%%%%%%%%%%%%%%%%%%

The forthcoming exploration of SD with unprecedented precision naturally 
invites to test the predictions of inflationary models, such as those based 
on axion monodromy, defined in section~\ref{sec:framework}.
We have determined in these constructions the observational features of 
$\mu$ and $y$ SD. By varying the different parameters and subjecting them 
to Planck constraints, we identified in section~\ref{sec:sdam} some values 
that maximize the resulting distortions for some benchmark models (with the 
powers $p\in\{\nicefrac23,1,\nicefrac43\}$ of the monomial contribution to 
the inflationary potential), see tables~\ref{table:sdam} and~\ref{tab:ModelParameters2}. 
Our main results are displayed in figures~\ref{plot:axm2}--\ref{plot:axm5}. 
Interestingly, accepting the possibility of a drifting axion decay parameter 
$p_f$ that varies during inflation, the resulting distortions exhibit a 
wave-damping behavior, which may be observable. If the drifting does not 
vary and develops a value $p_f\lesssim-0.7$, the associated SD become sizable. 
Beyond this feature, we find that the distortions arising from axion monodromy 
are distinguishable from the most conservative SD signal based on current 
$\Lambda$CDM observations, with up to 10\% deviations with respect to standard 
values in the observable frequency window, cf.\ figure~\ref{plot:axm5}.  

On a less positive note, we find it challenging for future missions, such as 
PIXIE and Super-PIXIE, to discriminate between inflationary models based on a 
power-law potential and axion monodromy. SD in axion monodromy with 
$f_0 \gtrsim 4\times10^{-3}$ and $p_f \lesssim 0.2$ differ maximally from the 
signal of a standard scenario based on a power-law potential. However, the 
difference is just of order 1\% or less. Consequently, one needs greater 
experimental accuracy than currently achievable to notice such small discrepancies. 
We expect that this caveat shall be solved by PIXIE setups capable of improving 
the sensitivity by at least 100 times, such as those already proposed in~\cite{Fu:2020wkq}.

The various cosmological probes in function and under development are paramount 
to falsify the predictions of axion monodromy and other theoretical proposals 
for inflation and, hence, to test our knowledge of the early Universe. Thus, 
in particular, a full updated analysis of the SD in addition to other 
astrophysical and cosmological signals from these models must be carried out. 
The current work represents a step towards this goal. It would be interesting 
to also include extensions to the scenario studied here, such as those proposed 
e.g.\ in~\cite{Bhattacharya:2022fze,Ballesteros:2019hus}.

\subsection*{Acknowledgments}

RHO acknowledges the financial support from UNACH and ICTIECH. JM acknowledges the support 
by the program {\it Investigadores por M\'exico/C\'atedras CONACYT} through 
the project 802. SRS thanks Carlos Alvarez Segura for useful discussions.

{\small
%\bibliographystyle{OurBibTeX}
%\bibliography{monodromy_sd}
\providecommand{\bysame}{\leavevmode\hbox to3em{\hrulefill}\thinspace}

}

\end{document}